\documentclass[11pt]{article}

\usepackage{amsfonts,amssymb}
\usepackage{dsfont}
\usepackage{mathrsfs}
\usepackage{tikz}
\usepackage{pgfplots}
\usepackage{chet}

\usetikzlibrary{decorations.markings}

\tikzset{->/.style={decoration={markings,mark=at position 1 with
  {\arrow[scale=1.5]{>}}},postaction={decorate}}}

\newcommand{\tikzmark}[1]{\tikz[overlay,remember picture] \coordinate
(#1);}

\pgfplotsset{compat=1.5.1}

\pgfplotsset{
    standard/.style={
        axis x line=bottom,
        axis y line=left,
        every axis x label/.style={at={(current axis.right of
          origin)},anchor=west},
        every axis y label/.style={at={(current axis.above
          origin)},anchor=south}}}

\DeclareMathOperator{\Tra}{Tr}
\DeclareMathOperator{\tra}{tr}
\DeclareMathOperator{\STr}{STr}
\DeclareMathOperator{\FT}{F\!.\!T\!.}
\DeclareMathOperator{\LiTwo}{\text{Li}_2}

\newcommand{\vev}[1]{\langle #1 \rangle}

\newcommand{\tNc}{\widetilde{N}_c}
\newcommand{\tZ}{\widetilde{Z}}
\newcommand{\tq}{\tilde{q}}
\newcommand{\tvarphi}{\tilde{\varphi}}
\newcommand{\tQ}{\widetilde{Q}}
\newcommand{\tF}{\widetilde{F}}
\newcommand{\tchi}{\tilde{\chi}}
\newcommand{\trho}{\tilde{\rho}}
\newcommand{\<}{\langle}
\renewcommand{\>}{\rangle}
\newcommand{\cQ}{\mathcal{Q}}
\newcommand{\cS}{\mathcal{S}}
\newcommand{\cQb}{\bar{\mathcal{Q}}}
\newcommand{\cD}{\mathcal{D}}
\newcommand{\sD}{\mathscr{D}}
\newcommand{\cM}{\mathcal{M}}
\newcommand{\bbM}{\mathbb{M}}
\newcommand{\cG}{\mathcal{G}}

\preprint{CERN-PH-TH/2012-334\\
SU-ITP-12/40\\
UCSD-PTH-12-19}

\title{Field-theoretic Methods in Strongly-Coupled Models\\
of General Gauge Mediation}

\author{Jean-Fran\c{c}ois Fortin$^{\ast,\dag,}$\email{jean-francois.fortin@cern.ch} and Andreas Stergiou$^{\$,}$\email{stergiou@physics.ucsd.edu}}

\affiliation{$^{\ast}$Theory Division, Department of Physics, CERN, CH-1211 Geneva 23, Switzerland\\
$^\dag$Stanford Institute for Theoretical Physics, Department of Physics, Stanford University, Stanford, CA 94305, USA\\
$^{\$}$Department of Physics, University of California, San Diego, La Jolla, CA 92093, USA}

\abstract{An often-exploited feature of the operator product expansion
(OPE) is that it incorporates a splitting of ultraviolet and infrared
physics.  In this paper we use this feature of the OPE to perform simple,
approximate computations of soft masses in gauge-mediated supersymmetry
breaking.  The approximation amounts to truncating the OPEs for
hidden-sector current-current operator products. Our method yields
visible-sector superpartner spectra in terms of vacuum expectation values
of a few hidden-sector IR elementary fields.  We manage to obtain
reasonable approximations to soft masses, even when the hidden sector is
strongly coupled. We demonstrate our techniques in several examples,
including a new framework where supersymmetry-breaking arises both from a
hidden sector and dynamically.}

\date{December 2012} 

\begin{document}

\maketitle

\newsec{Introduction}
The theoretical appeal of supersymmetry (SUSY) makes imperative the study
of the phenomenology of its breaking.  The Large Hadron Collider (LHC) has
not yet found signs of low-scale SUSY, but abandoning SUSY at this early
stage in experimental discovery would be premature.  Nevertheless, SUSY
extensions of the Standard Model are now tightly constrained by
experimental data, and it appears that the simplest among them are not
likely to survive as viable candidates for phenomenology. Therefore, new
models of SUSY breaking as well as new tools for their analysis remain
useful in exploring physics beyond the Standard Model. It would of course
be ideal if tools were developed that could be used at strong coupling,
since if SUSY is a symmetry of nature at some high scale, then it may very
well reside in a model that is strongly-coupled at low energies.

In the context of gauge mediation of SUSY breaking (for a review
see~\cite{Giudice:1998bp}) a formalism exists, known as general gauge
mediation (GGM), that allows one to study such models in a unified
fashion~\cite{Meade:2008wd, Buican:2008ws, Dumitrescu:2010ha}. More
specifically, SUSY-breaking parameters in the minimal supersymmetric
standard model (MSSM) are generated in models of gauge-mediated SUSY
breaking via two-point correlators of gauge-current superfields of the
hidden, SUSY-breaking sector. This, then, dictates that a current analysis
is possible, and allows one to understand the generation of soft masses in
the MSSM Lagrangian.

Such an analysis benefits strongly from the use of the operator product
expansion (OPE). In $\mathcal N=1$ superconformal theories OPEs of current
correlators were studied in~\cite{Fortin:2011nq}, where the superconformal
symmetry was seen, as expected, to relate the OPEs of different components
of the gauge-current superfield. Of course the study of the OPE is
motivated by the fact that the OPE is one of the few tools that allows us
to extract useful information even at strong coupling. This is reflected in
the wealth of applications of the OPE in QCD.

The results of~\cite{Fortin:2011nq} were applied to the case of GGM
correlators in~\cite{Fortin:2011ad}. Part of the motivation for that work
was the observation that, even in theories that break the superconformal
symmetry explicitly, one can introduce spurions to render the breaking
spontaneous.  The spurions are fully dynamical in the ultraviolet (UV), and
an OPE analysis can be carried out to determine Wilson coefficients of
spurionic operators in operator products. In the infrared (IR) the spurions
acquire vacuum expectation values (vevs), and the Wilson coefficients have
to be evolved from the UV according to their renormalization-group
equation. It was shown in~\cite{Fortin:2011ad} that, in the case of minimal
gauge mediation (MGM), soft masses could be approximated very well by only
the leading spurionic term in the current-current OPE that develops a
SUSY-breaking vev.

In MGM one can actually compute the full gaugino and sfermion masses using
the OPE~\cite{Fortin:2011ad}. This is a rather special case and one cannot
typically expect to be able to compute the complete current-current OPE.
Nevertheless, it is physically acceptable to truncate the OPE and carry out
the calculation of the soft masses, since the truncation is not expected to
alter significantly the essential results. The error introduced in
truncating the OPE allows only an approximate determination of the soft
masses, up to $\mathcal O(1)$ overall factors which may be unimportant.

The technology developed in~\cite{Fortin:2011ad} may be used in
strongly-coupled models of SUSY breaking. This is because the determination
of Wilson coefficients is done in the UV, where asymptotic freedom allows
for a perturbative computation, while non-perturbative effects are
contained in the vevs of operators, i.e.\ are captured by IR quantities.
Thus, at least at the qualitative level, one is able to use the methods
of~\cite{Fortin:2011ad} in order to understand the generation of soft
masses in the MSSM, even when the SUSY-breaking sector is strongly-coupled
in the IR.  In theories where weakly-coupled duals exist, it is also possible
to check the strongly-coupled computations at the quantitative level by
comparing results obtained with both methods.  As we will see the
approximations discussed here are indeed reasonable up to factors of order
one, suggesting that relevant information can be extracted from them even in
the strong-coupling regime.

Of course AdS/CFT~\cite{Maldacena:1997re} is another tool one can use in
order to understand the behavior of field theories at strong coupling.
Indeed, the realization of GGM in holography has been considered by
numerous authors, see e.g.~\cite{Benini:2009ff, McGuirk:2009am,
McGarrie:2010kh, McGarrie:2010yk, McGuirk:2011yg, Skenderis:2012bs,
Argurio:2012cd, McGarrie:2012ks, Argurio:2012bi, McGarrie:2012fi}. The main
theme of these works is the description of GGM correlators by holographic
methods. In this paper, however, our methods will be strictly
field-theoretic and four-dimensional.

$\mathcal N=1$ supersymmetric QCD (SQCD) is an ideal candidate for the
application of our methods. In the free magnetic range of the massive
theory Intriligator, Seiberg and Shih (ISS) demonstrated the existence of a
metastable SUSY-breaking vacuum~\cite{Intriligator:2006dd}. In their
treatment they used the power of Seiberg duality~\cite{Seiberg:1994pq} in
order to establish their result in the strongly-coupled regime of the
electric theory.  The big global symmetry of SQCD in the ISS vacuum allows
its use as the hidden SUSY breaking sector in the context of gauge
mediation.  Phenomenologically, however, there is a problem due to an
accidental R-symmetry which precludes Majorana masses for the gauginos.

Although modifications of the ISS scenario have been proposed in the
literature, see e.g.~\cite{Amariti:2006vk, Kitano:2006xg, Haba:2007rj,
Giveon:2007ew, Xu:2007az, Zur:2008zg, Giveon:2008ne, Franco:2009wf,
Barnard:2009ir, Bertolini:2011hy, Amariti:2012pg, Essig:2008kz}, in this
paper we consider a new deformation where we add an additional spontaneous
breaking of SUSY from a singlet chiral superfield.  This superfield
acquires its vev through its own dynamics, about which we will remain
agnostic.  This new model is similar to MGM but with messengers strongly
interacting through another gauge group.  As we will see, with this
deformation our theory develops ISS-like vacua but with a broken
R-symmetry.  In our example there are no SUSY vacua anywhere in field
space, but the ISS-like vacua we find should be metastable against decay to
other SUSY-breaking vacua with lower energy.

The paper is organized as follows. In section \ref{review} we review
background material related to our work. We give a lightning review of
$\mathcal N=1$ SQCD, as well as a quick overview of gauge mediation, GGM,
and the role of the OPE in our considerations. In section \ref{SQCDSUSYbr}
we present the analysis of our deformation of ISS. We also recover MGM and
pure ISS as limits of our deformed SQCD. We conclude in section \ref{conc}.
Appendix \ref{app:Supspec} contains weakly-coupled computations of the
superpartner spectrum for general messenger sectors.  We use notation and
conventions of Wess \& Bagger~\cite{Wess:1992cp}.

\newsec{\texorpdfstring{$\mathbf{\mathcal{N}=1}$}{N=1} SQCD, gauge
mediation of SUSY breaking, and the OPE}[review]
In this section we first review the aspects of $\mathcal{N}=1$ SQCD and
gauge mediation which are necessary for our purposes.  This section is far
from self-contained and the reader is referred to the literature,
e.g.~\cite{Intriligator:1995au}, for completeness.

\subsec{Essentials of \texorpdfstring{$\mathcal{N}=1$}{N=1} SQCD}
SQCD with $N_c$ colors and $N_f$ flavors is an $\mathcal{N}=1$
supersymmetric $ SU(N_c)$ gauge theory with $N_f$ quark flavors $Q^i$
(left-handed quarks) which are chiral superfields transforming in the
$\mathbf{N_c}$ of $ SU(N_c)$, and $N_f$ quark flavors
${\widetilde{Q}}_{\tilde{\imath}}$ (left-handed antiquarks) which are
chiral superfields transforming in the $\mathbf{\overline{N}_c}$ of $
SU(N_c)$, where $i,\tilde{\imath}=1,\ldots,N_f$ are flavor
indices.\foot{Note that there are no Fayet--Iliopoulos terms since the
gauge group does not contain $U(1)$ factors.}

There is a large global symmetry in SQCD---the relevant representations and
charge assignments are shown in Table \ref{SQCDAnom}.
\begin{table}[ht]
  \centering
  \begin{tabular}{c|ccccc}
     &  $SU(N_f)_\text{L}$ &  $SU(N_f)_\text{R}$ &  $U(1)_\text{B}$ &
     $U(1)_\text{A}$
     &  $U(1)_{\text{R}^\prime}$ \\
     \hline
     $Q$  & $\mathbf{N_f}$ & $\mathbf{1}$ & $1$ & $1$ & $1$ \\
     $\widetilde{Q}$ & $\mathbf{1}$ & $\mathbf{\overline{N}_f}$ & $-1$
     & $1$ & $1$
  \end{tabular}
  \caption{Matter content of SQCD and its (anomalous) transformation
  properties.}
  \label{SQCDAnom}
\end{table}

However, the $ U(1)_\text{A}\times U(1)_{\text{R}^\prime}$ symmetry is
anomalous. A single $U(1)$ R-symmetry, which we will denote $
U(1)_{\text{R}}$, survives and is a full quantum symmetry.  Thus, the
global symmetry of the quantum theory is $ SU(N_f)_\text{L}\times
SU(N_f)_\text{R}\times U(1)_\text{B}\times U(1)_\text{R}$ with the
appropriate R-charge assignment as shown in Table \ref{SQCDNonAnom}.
\begin{table}[ht]
  \centering
  \begin{tabular}{c|cccc}
    &  $SU(N_f)_\text{L}$ &  $SU(N_f)_\text{R}$ &  $U(1)_\text{B}$ &
     $U(1)_\text{R}$ \\
     \hline
     $Q$  & $\mathbf{N_f}$ & $\mathbf{1}$ & $1$ & $1-N_c/N_f$ \\
     $\widetilde{Q}$ & $\mathbf{1}$ &$\mathbf{\overline{N}_f}$ & $-1$ &
     $1-N_c/N_f$
  \end{tabular}
  \caption{Matter content of SQCD and its (non-anomalous) transformation
  properties.} \label{SQCDNonAnom}
\end{table}

To make our notation more convenient we define the matrices\\
\eqn{Q =\,\,
  \begin{pmatrix}
    \begin{pmatrix}
      \tikzmark{Qleftcorner}\phantom{Q}\\
      Q^1\\
      \phantom{Q}\\
    \end{pmatrix} & \cdots\xspace &
    \begin{pmatrix}
      \phantom{Q}\\
      Q^{N_f}\\
      \phantom{Q}
    \end{pmatrix}
  \end{pmatrix},\qquad\qquad
  \widetilde{Q} =\,\,
  \begin{pmatrix}
    (\begin{matrix}
      \tikzmark{tildeQleftcorner}\phantom{Q} & \,\,\widetilde
      Q^{1}\,\, & \phantom{Q}
    \end{matrix}) \\
    \vdots\xspace \\
    (\begin{matrix}
      \phantom{Q} & \widetilde Q^{N_f} & \phantom{Q}
    \end{matrix})
  \end{pmatrix},
  \tikz[overlay,remember picture] {
  \draw[->] ([xshift=-1ex,yshift=4ex]Qleftcorner)--+(0.55,0)
  node[xshift=-1.2ex,at start] {$i$};
  \draw[->] ([xshift=-5ex,yshift=0.5ex]Qleftcorner)--+(0,-0.57)
  node[yshift=1.5ex,at start] {$a$};
  \draw[->] ([xshift=0.8ex,yshift=3.5ex]tildeQleftcorner)--+(0.55,0)
  node[xshift=-1.3ex,at start] {$a$};
  \draw[->] ([xshift=-3.3ex]tildeQleftcorner)--+(0,-0.55) node[yshift=1.7ex,at
  start] {$\tilde\imath$};
}}
where $a=1,\ldots,N_c$ is a fundamental or antifundamental color index. In
this notation the Lagrangian of SQCD is\foot{$\Tra$ denotes a sum over both
fundamental gauge and flavor indices, while $\tra$ denotes a sum over
adjoint gauge indices only, e.g.\
\eqn{\Tra Q^\dagger T^I Q \equiv Q^\dagger_{ib}(T^I)^b_c Q^{ic}\qquad
\text{and}\qquad\tra W^\alpha W_\alpha\equiv W^{\alpha I}W_{\alpha}^I.}
}
\eqna{\mathscr{L}_\text{SCQD} &= \int d^4\theta\, \Tra(Q^\dagger e^{2gV} Q
+\tQ e^{-2gV}\tQ^\dagger)+\left(\int d^2\theta\, \tra W^\alpha
W_\alpha+\text{h.c.}\right).}
In components (and after integrating out the auxiliary fields), this
becomes
\eqna{\mathscr{L}_\text{SCQD} &= -\tra(\tfrac14 F_{\mu\nu}F^{\mu\nu}
+ i\lambda\sigma^\mu\sD_\mu\bar{\lambda})
-\Tra[\sD_\mu Q^\dagger\sD^\mu Q +\sD_\mu\tQ\sD^\mu\tQ^\dagger
+i\bar{\psi}\bar{\sigma}^\mu\sD_\mu\psi + i\tilde{\psi}\bar{\sigma}^\mu
\sD_\mu \bar{\tilde{\psi}}\\
&\quad -i\sqrt2
g(Q^\dagger\lambda\psi-\bar{\psi}\bar{\lambda}Q-\tilde{\psi}\lambda
\tQ^\dagger + \tQ\bar{\lambda}\bar{\tilde{\psi}})] -
\tfrac{1}{2}g^2\sum_{I=1}^{N_c^2-1} [\Tra(Q^\dagger T^I Q -
{\widetilde{Q}} T^I {\widetilde{Q}}^\dagger)]^2,}
where $\sD_\mu=\partial_\mu+igA_\mu^I T^I(R)$ is the gauge-covariant
derivative.  Note that SQCD only has D-term contributions to the scalar
potential,
\eqn{\mathscr{V}_\text{SCQD} = \tfrac{1}{2}g^2\sum_{I=1}^{N_c^2-1}
[\Tra(Q^\dagger T^I Q - {\widetilde{Q}} T^I {\widetilde{Q}}^\dagger)]^2,}
where $T^I$ are $SU(N_c)$ generators with $I=1,\ldots,N_c^2-1$ the adjoint
color index.  This scalar potential has a large vacuum degeneracy, which is
however lifted when masses for the quarks are added.

\subsubsec{Masses for the flavors}
The lowest-dimensional gauge-invariant chiral superfield one can construct
from $Q^i$ and ${\widetilde{Q}}_{\tilde{\imath}}$, namely the mesonic
superfield\foot{$\Tra(\cdot)_{(x,y)}$ denotes a sum over color indices up
to $x$ and flavor indices up to $y$.  Hence,
$\Tra(\cdot)\equiv\Tra(\cdot)_{(N_c,N_f)}$.}
\eqn{M^i_{\tilde{\imath}} = \Tra({\widetilde{Q}}_{\tilde{\imath}}
Q^i)_{(N_c,0)},}
can be used to give gauge-invariant masses to all quark flavors.  The
Lagrangian of massive SQCD (mSQCD) is then
\eqn{\mathscr{L}_\text{mSCQD}=\mathscr{L}_\text{SCQD}+\left(\int d^2\theta
\, W_\text{tree}+\text{h.c.}\right)\!,}
where $W_\text{tree}=\Tra(mM)_{(0,N_f)}$, with $m$ a nondegenerate
$N_f\times N_f$ mass matrix.  Note that the inclusion of masses breaks the
non-Abelian part of the global symmetry to one of its subgroups. The scalar
potential in $\mathscr{L}_\text{mSCQD}$ is
\eqn{\mathscr{V}_\text{mSCQD} =\Tra(m m^\dagger Q^\dagger Q +
m^\dagger m\widetilde{Q}{\widetilde{Q}}^\dagger)+
\tfrac{1}{2}g^2\sum_{I=1}^{N_c^2-1} [\Tra(Q^\dagger T^I Q -
\widetilde{Q} T^I {\widetilde{Q}}^\dagger)]^2,}
and includes the anticipated mass terms. The vacuum degeneracy of
$\mathscr{V}_\text{SCQD}$ is lifted in $\mathscr{V}_\text{mSCQD}$ due to
the mass terms.

\subsec{Essentials of gauge mediation}
Mediation of SUSY breaking was born to address phenomenological impasses
reached by trying to break SUSY within the observable sector of
supersymmetric extensions of the standard model. As an example, supertrace
conditions that remain even after SUSY is broken are hard to satisfy
consistently with the observed low-mass spectrum of particles
\cite{Dimopoulos:1981zb}.

Gauge mediation requires that SUSY be broken in a hidden sector with the
breaking communicated to the MSSM through the familiar gauge interactions,
thus avoiding new sources of flavor-changing neutral currents, a generic
problem in models of gravity-mediated SUSY breaking.  All soft
SUSY-breaking terms in the MSSM Lagrangian are generated via loop effects,
and desired phenomenology is obtained very naturally, except, of course,
for the notorious $\mu/B_\mu$ problem~\cite{Dvali:1996cu}. For an extensive
review of theories with gauge mediation the reader is referred
to~\cite{Giudice:1998bp}.

In the minimal incarnation of gauge mediation one assumes the existence of
a hidden sector that contains a gauge singlet chiral superfield $S$, as
well as a messenger sector with fields $\Phi,\widetilde{\Phi}$ in complete
GUT representations so that gauge-coupling unification is not spoiled.
Through interactions in the hidden sector $S$ develops a vev both in its
first and its last component, $\vev{S}=\vev{S}+\theta^2\vev{F_S}$. The
superpotential that couples the hidden sector with the messenger sector is
$W_{\text{h}\otimes\text{m}}\propto S\Tra(\widetilde{\Phi}\Phi)$, such that
the SUSY breaking of the hidden sector is fed into the messenger sector.
The usual gauge interactions then communicate the SUSY breaking to the
supersymmetric extension of the standard model generating the appropriate
soft SUSY-breaking terms.

\subsubsec{General gauge mediation}
A unified and powerful framework for the study of gauge mediation, dubbed
general gauge mediation, was developed in~\cite{Meade:2008wd, Buican:2008ws,
Dumitrescu:2010ha}. In GGM soft terms are written in terms of one- and
two-point correlators of components of a current (linear) superfield of the
hidden sector,
\eqn{\mathcal{J}(z)=J(x)+i\theta j(x)-i\bar\theta \bar\jmath(x)
-\theta \sigma ^\mu \bar \theta j_\mu(x) +\cdots,}[Jz]
where the ellipsis stands for derivative terms, following from the
conservation equations $\cD^2\mathcal{J}=\bar{\cD}^2\mathcal{J}=0$.\foot{In
this paper $D$ is the D-term, thus we use $\cD$ for the covariant
derivatives.} Among the virtues of GGM is its ability to disentangle
genuine characteristics of gauge mediation from possible model-dependent
features.  GGM also leads to phenomenological superpartner-mass sum rules
that, if verified by the LHC, will identify gauge mediation as the dominant
means by which SUSY is broken in nature (see e.g.~\cite{Jaeckel:2011ma,
Jaeckel:2011qj} for a renormalization group study of the above-mentioned
sum rules).  Moreover, GGM encompasses strongly-coupled hidden sectors at
the qualitative level and also at the quantitative level, at least in
principle.  In our view this is the greatest strength of GGM, which is
nevertheless largely unexplored. In the next section it will be discussed
extensively.

The correlators one considers in GGM are (using the conventions
of~\cite{Dumitrescu:2010ha})
\eqna{
\<J(x)J(0)\> & =C_0(x) \xrightarrow{\FT}\widetilde C_0(p),\\
\<j_{\alpha}(x)\bar{\jmath}_{\dot\alpha}(0)\> &
=-i\sigma^\mu_{\alpha\dot{\alpha}}\partial_\mu C_{1/2}(x)
\xrightarrow{\FT}\sigma ^\mu _{\alpha \dot \alpha
}p_\mu \widetilde C_{1/2}(p),\\
\<j_\mu(x)j_\nu(0)\> & =(\eta_{\mu\nu}\partial^2
-\partial_\mu\partial_\nu)C_1(x) \xrightarrow{\FT}
-(\eta_{\mu\nu}p^2-p_\mu p_\nu)\widetilde C_1(p) ,\\
\<j_\alpha(x)j_\beta(0)\> & =\epsilon_{\alpha\beta}B_{1/2}(x)
\xrightarrow{\FT}\epsilon_{\alpha\beta}\widetilde B_{1/2}(p),\\
}[correls]
where $\FT$ stands for Fourier-transforming, $\FT\equiv i\int d^4 x\,
e^{-ip\cdot x}$.  It was realized in~\cite{Buican:2008ws} that for the soft
masses, for example, only the one-point function $\<J(x)\>$ and the
correlator $\<J(x)J(0)\>$ are needed:\foot{Since $Q$ is used in this paper
for the quarks of SQCD, we use $\cQ$ to denote the SUSY generator.  $\cQ$
always acts with an adjoint action, e.g.\ $\cQ^2(\mathcal{O}(x))\equiv
\{\cQ^\alpha,[\cQ_\alpha,\mathcal{O}(x)]\}$.}
\eqna{
M_{\text{gaugino}} & =\frac{i\pi\alpha_\text{SM}}{d(G)}
\int d^4x\,\<\cQ^2(J^A(x)J^A(0))\>,\\
m_{\text{sfermion}}^2 & =4\pi
Y\alpha_\text{SM}\<J(x)\>+\frac{iC_2(R)\alpha_\text{SM}^2}{8d(G)}\int
d^4x\ln(x^2M_\text{m}^2)
\<\cQb^2\cQ^2(J^A(x)J^A(0))\>,
}[GGMsoft]
where $M_\text{m}$ is a supersymmetric scale in the hidden-sector theory,
e.g.\ the messenger scale. For clarity, the appropriate MSSM gauge group
index $A$ has been reintroduced.\foot{The MSSM gauge group is chosen to be
a GUT $SU(N)$ subgroup of the hidden-sector global symmetry group where
$A=1,\ldots,N^2-1$ is the appropriate adjoint index.}

Using the results of~\cite{Osborn:1998qu} it was pointed out
in~\cite{Fortin:2011nq} that, within a superconformal field theory, the
superconformal algebra and current conservation are powerful enough to
relate all possible two-operator products of components of the current
superfield \Jz to the operator product $J(x)J(0)$. Consequently, only the
correlator $\<J(x)J(0)\>$ is necessary, while all other correlators in
\correls can be expressed in terms of $\<J(x)J(0)\>$ with the help of the
superconformal group. From~\cite{Fortin:2011nq} one has
\eqna{
j_\alpha (x) j_\beta (0)&=\frac{1}{x^2}\cQ_\beta (ix\cdot \sigma \bar \cS)_\alpha
(J(x)J(0)),\displaybreak[0]\\
j_\alpha(x)\bar{\jmath}_{\dot{\alpha}}(0) &
=\frac{1}{x^4}\left[(\cS\,ix\cdot\sigma)_{\dot{\alpha}}(ix\cdot\sigma\bar{\cS})_\alpha-x^2\bar{\cQ}_{\dot{\alpha}}(ix\cdot\sigma\bar{\cS})_\alpha+2\Delta
_Jx^2(ix\cdot\sigma)_{\alpha\dot{\alpha}}\right](J(x)J(0)),\displaybreak[0]\\
j_\mu(x)j_\nu(0) & =\frac{1}{16x^8}\left[(x^2\eta_{\mu\rho}-2x_\mu
x_\rho)(\cS\sigma^\rho\bar{\cS}-\bar{\cS}\sigma^\rho \cS)\right.\displaybreak[0]\\
 &
 \quad\left.\quad\quad\quad\quad\quad\quad\quad\times\{x^4(\bar{\cQ}\bar{\sigma}_\nu
 \cQ-\cQ\sigma_\nu\bar{\cQ})+(x^2\eta_{\nu\lambda}-2x_\nu
 x_\lambda)(\cS\sigma^\lambda\bar{\cS}-\bar{\cS}\bar{\sigma}^\lambda
 \cS)\right.\displaybreak[0]\\
 &
 \quad\left.\quad\quad\quad\quad\quad\quad\quad\quad\quad\quad\quad\quad\quad
 \quad\quad\quad\quad\quad\quad
 -2x^2\left(\cQ\sigma_\nu\,ix\cdot\bar{\sigma}\cS-\bar{\cQ}\bar{\sigma}_\nu\,
 ix\cdot\sigma\bar{\cS}\right)\}\right.\displaybreak[0]\\
 &
 \quad\left.\quad\quad\quad-8i(\Delta_J+1)x^2(\eta_{\mu\nu}\eta_{\lambda\rho}-\eta_{\mu\lambda}\eta_{\nu\rho}-\eta_{\mu\rho}\eta_{\nu\lambda}-i\epsilon_{\mu\nu\lambda\rho})x^\lambda\right.\displaybreak[0]\\
 &
 \quad\left.\quad\quad\quad\quad\quad\quad\quad\times\{(x^2\eta^{\rho\delta}-2x^\rho
 x^\delta)\cS\sigma_\delta\bar{\cS}+x^2\bar{\cQ}\bar{\sigma}^\rho\,ix\cdot\sigma\bar{\cS}+4i\Delta_Jx^2x^\rho\}\right.\displaybreak[0]\\
 &
 \quad\left.\quad\quad\quad-8i(\Delta_J+1)x^2(\eta_{\mu\nu}\eta_{\lambda\rho}-\eta_{\mu\lambda}\eta_{\nu\rho}-\eta_{\mu\rho}\eta_{\nu\lambda}+i\epsilon_{\mu\nu\lambda\rho})x^\lambda\right.\displaybreak[0]\\
 &
 \quad\left.\quad\quad\quad\quad\quad\quad\quad\times\{(x^2\eta^{\rho\delta}-2x^\rho
 x^\delta)\bar{\cS}\bar{\sigma}_\delta
 \cS+x^2\cQ\sigma^\rho\,ix\cdot\bar{\sigma}\cS
 +4i\Delta_Jx^2x^\rho\}\right.\displaybreak[0]\\
 &
 \quad\left.\quad\quad\quad+32x^4\Delta_J(\Delta_J+1)(x^2\eta_{\mu\nu}-2x_\mu
 x_\nu)\right](J(x)J(0)),}
with $\cS,\bar{\cS}$ the superconformal supercharges. Implications of this
observation in the case of a UV asymptotically-free hidden sector (i.e.\
with approximate superconformal symmetry) and in particular in the example
of MGM were analyzed using the OPE in~\cite{Fortin:2011ad}, and we will rely
heavily here on the results of that paper. It is important to note that using
the OPE in the equations above and Fourier-transforming the results allow a
simple evaluation of the total cross-sections of the visible sector to the
hidden sector, with different mediators corresponding to the different
components of the MSSM vector superfields.  This is reminiscent of
electron-positron scattering to hadrons in QCD. In the following we
will focus on the superpartner spectrum, and will not discuss such
cross-sections.

As shown in~\cite{Fortin:2011ad} a complete expansion of \GGMsoft can be
obtained with the help of the $J(x)J(0)$ OPE which thus gives an
approximation to the soft MSSM SUSY-breaking masses even for
strongly-coupled hidden sectors.  The expansion relies on several
approximations (e.g.\ cuts at supersymmetric threshold, uniform convergence
of the OPE) but, at least in the simple case of MGM, a complete knowledge
of the OPE leads to an exact evaluation of the soft SUSY-breaking masses,
after analytic continuation of the sums.  To avoid complications such as
arduous OPE computations and analytic continuations, a further
approximation to \GGMsoft, given by
\eqna{
M_{\text{gaugino}} & \approx-\frac{\pi w^{AA}\alpha_\text{SM}}
{8d(G)M_\text{m}^2}\gamma_{Ki}\<\cQ^2(\mathcal{O}_i(0))\>,\\
m_{\text{sfermion}}^2 & \approx4\pi Y\alpha_\text{SM}\<J(x)\>
+\frac{C_2(R)w^{AA}\alpha_\text{SM}^2
}{64d(G)M_\text{m}^2}\gamma_{Ki}\<\cQb^2\cQ^2(\mathcal{O}_i(0))\>,
}[SoftMasses]
was introduced in~\cite{Fortin:2011ad}.  Here $w^{AB}$ is the OPE
coefficient of a scalar operator $K$ with classical scaling dimension 2 in
the OPE of two conserved currents (like, e.g. the Konishi current in MGM),
and $\gamma$ is the anomalous-dimension matrix of $K$ (see \eqref{OPE},
\eqref{AnomMatWeak} and \eqref{AnomMatStrong}). So, to get an approximation
to the soft MSSM SUSY-breaking masses, \textit{even in a theory with a
strongly-coupled hidden sector}, one only needs to identify the
lowest-dimension operators that have non-zero vevs after acted upon with
$\cQ^2$ and $\cQb^2\cQ^2$.

In the example of MGM there is only one such operator, namely $S^\dagger
S$, and calculating its mixing with the Konishi current one finds that the
approximation to the soft masses \SoftMasses is actually only a factor of 2
smaller than the usually quoted answers~\cite{Martin:1996zb}.  For more
details the reader is referred to section \ref{MGM}
and~\cite{Fortin:2011ad}.

\newsec{SQCD as the SUSY-breaking sector}[SQCDSUSYbr]
To be specific, in this paper we take the messenger sector of gauge
mediation to be SQCD without masses for the quarks but, instead, with
K\"{a}hler potential and superpotential for matter fields given by
\eqna{K_\text{e}&=\Tra (Q^\dagger Q+
\widetilde{Q}\widetilde{Q}^\dagger)+S^\dagger S,\\
W_\text{e}&=\xi S\Tra \widetilde{Q}Q,}[Elec]
where $S$ is the MGM-like singlet field which has non-vanishing vacuum
expectation value $\vev{S}=\vev{S}+\theta^2\vev{F_S}$, and
$Q,\widetilde{Q}$ are the messenger fields which are $N_f$ flavors of
$SU(N_c)$ fundamental and antifundamental superfields.  The non-Abelian
part of the global symmetry of SQCD is thus broken to its diagonal
subgroup, $SU(N_f)_\text{L}\times SU(N_f)_\text{R}\to SU(N_f)_\text{V}$,
which contains $SU(N)$, a grand-unified extension of the MSSM gauge group.
The coupling $\xi$ is assumed weak.  We will refer to SQCD with an extra
singlet and the superpotential \Elec as sSQCD.  We stress that it is
straightforward to repeat the analysis for more general messenger sectors
and hidden sectors.

In order to use the approximation \SoftMasses in this framework, it is
necessary to determine the $J(x)J(0)$ OPE at the lowest non-trivial order
as well as the appropriate anomalous dimension matrix.

Note that non-perturbative effects (instantons) contribute both to the vevs
of operators appearing on the right-hand side of the OPE and to the
(perturbative) OPE coefficients themselves~\cite{Novikov:1980uj}.
Furthermore, for operator products satisfying the chirality selection rule,
instantons can lead to new non-perturbative contributions on the right-hand
side of the OPE, i.e.\ operators with purely non-perturbative OPE
coefficients~\cite{Amati:1984uz}.  Instanton corrections of the first type
do not modify the OPE coefficients at lowest order and are thus
non-negligible only for vevs of operators.  Instanton corrections of the
second type lead to new non-perturbative OPE contributions which can
dominate over the perturbative ones.\foot{It is important to notice that
both types of non-perturbative contributions to the OPE coefficients are
calculable.  Thus, as usual, the OPE coefficients are fully calculable,
while all incalculable non-perturbative effects are contained in the vevs
of operators.}  Since the $J(x)J(0)$ OPE is non-trivial at the classical
level and does not satisfy the chirality selection rule, for our purposes
non-perturbative contributions that are calculable can be safely ignored.

The currents of interest for the evaluation of the $J(x)J(0)$ OPE are
\eqna{J^A&=\Tra(Qt^AQ^\dagger-\widetilde{Q}^\dagger
t^A\widetilde{Q})_{(N_c,N)},\\
K&=\Tra(Q^\dagger Q+\widetilde{Q}\widetilde{Q}^\dagger)_{(N_c,N)},}
where we denote the $SU(N)$ generators by $t^A$ to avoid confusion with the
$SU(N_c)$ generators $T^I$.  At the classical level the OPE is simply
\eqn{J^A(x)J^B(0)=\frac{N_c\delta^{AB}}{16\pi^4x^4}\mathds{1}
+\frac{w^{AB}}{4\pi^2x^2}K(0)+\cdots,}[OPE]
where $w^{AB}=\delta^{AB}/N$, while the one-loop anomalous-dimension matrix
between $K$ and $S^\dagger S$ is
\eqn{\gamma=\begin{pmatrix}
\gamma_{K,K} & \gamma_{K,S^\dagger S}\\
\gamma_{S^\dagger S,K} & \gamma_{S^\dagger S,S^\dagger S}
\end{pmatrix}\xrightarrow[\text{coupling}]{\text{weak}}\frac{1}{8\pi^2}
\begin{pmatrix}
2C_2(N_c)g^2 & 2NN_c|\xi|^2\\
|\xi|^2 & 0
\end{pmatrix}.}[AnomMatWeak]
Note here that although computable in the weak-coupling regime, the
anomalous dimensions are large in the IR for strongly-coupled theories and
are therefore kept undetermined in the following.  The soft SUSY-breaking
masses are
\eqna{
M_{\text{gaugino}} & \approx-\frac{\pi\alpha_\text{SM}}{8N|\xi\vev{S}|^2}
\left[\gamma_{K,K}\<\cQ^2(K)\>+\gamma_{K,S^\dagger S}\<\cQ^2(S^\dagger S)\>\right],\\
m_{\text{sfermion}}^2 & \approx\frac{C_2(R)\alpha_\text{SM}^2}
{64N|\xi\vev{S}|^2}\left[\gamma_{K,K}\<\cQb^2\cQ^2(K)\>
+\gamma_{K,S^\dagger S}\<\cQb^2\cQ^2(S^\dagger S)\>\right],
}[MassApprox]
since the supersymmetric mass scale $M_\text{m}=|\xi\vev{S}|$ and $\<J\>=0$
for a non-Abelian group.

These expressions can be further simplified using the supersymmetry algebra
and the Konishi anomaly~\cite{Konishi:1983hf} (in Wess--Zumino gauge) in
the $\alpha_\text{SM}\to0$ limit:
\eqna{\cQ^2(S^\dagger S) &=4S^\dagger F_S,\displaybreak[0]\\
\cQb^2\cQ^2(S^\dagger S) &= 16(F_S^\dagger F_S^{\phantom{\dagger}}
-i\bar{\psi}_S\bar{\sigma}^\mu\partial_\mu\psi_S + S^\dagger\partial^2S),\displaybreak[0]\\
\cQ^2(K) &=4\left[\Tra(Q^\dagger F+\tF\tQ^\dagger)_{(N_c,N)}
+\frac{Ng^2}{16\pi^2}\tra\bar{\lambda}\bar{\lambda}\right],\displaybreak[0]\\
\cQb^2\cQ^2(K) &=16\left[\vphantom{\left|\frac{\mu_F^3}{2h\mu^3}\right|}
\Tra(F^\dagger F-i\bar{\psi}\bar{\sigma}^\mu\sD_\mu\psi
+Q^\dagger\sD^2Q+i\sqrt{2}g(Q^\dagger\lambda\psi-\bar{\psi}
\bar{\lambda}Q)+gQ^\dagger DQ)_{(N_c,N)}\right.\\
&\left.\quad\hspace{0.6cm}+\{(Q,\psi,F,g)\to(\tQ,\tilde{\psi},\tF,-g)\}-
 \frac{Ng^2}{32\pi^2}\tra(2DD-4i\lambda\sigma^\mu
 \sD_\mu\bar{\lambda}-F_{\mu\nu}F^{\mu\nu})\right].}
Note that $\cQb^2\cQ^2(S^\dagger S, K)$ are real up to total derivatives.
After using the equations of motion (we omit the ones for the fields with a
tilde),
\begin{gather*}
F = -\xi^* S^\dagger \tQ^\dagger,\qquad\qquad
D^I = -g\Tra(Q^\dagger T^IQ-\tQ T^I\tQ^\dagger),\\
\sD^2 Q = -i\sqrt2 g\lambda \psi - gDQ + |\xi|^2S^\dagger S Q -
\xi^*F_S^\dagger \tQ^\dagger,\\
i\bar{\sigma}^\mu\sD_\mu\psi = -i\sqrt2 g\bar{\lambda}Q-\xi^*S^\dagger
\bar{\tilde{\psi}},\qquad\qquad
i\sigma^\mu\sD_\mu\bar{\lambda}^I = i\sqrt2 g(Q^\dagger T^I\psi
-\tilde{\psi}T^I\tQ^\dagger),
\end{gather*}
the approximations \MassApprox can be written in terms of vacuum
condensates of UV elementary fields as
\eqna{
M_{\text{gaugino}} & \approx\frac{\pi\alpha_\text{SM}}{2N|\xi\vev{S}|^2}
\left[2\xi^*\gamma_{K,K}\<S^\dagger\Tra(Q^\dagger\tQ^\dagger)
_{(N_c,N)}\>-\frac{Ng^2}{16\pi^2}\gamma_{K,K}\<\tra\bar{\lambda}
\bar{\lambda}\>-\gamma_{K,S^\dagger S}\<S^\dagger F_S\>\right]\\
m_{\text{sfermion}}^2 & \approx\frac{C_2(R)\alpha_\text{SM}^2}{4N|\xi
\vev{S}|^2}\left[2\xi^*\gamma_{K,K}\<\xi S^\dagger S K+
\Tra(S^\dagger\bar{\psi}\bar{\tilde{\psi}} - F_S^\dagger
Q^\dagger\tQ^\dagger)_{(N_c,N)}\>\right.\\
&\hspace{4cm}\left.-\frac{Ng^2}{32\pi^2}\gamma_{K,K}\<\tra(2DD-4E
-F_{\mu\nu}F^{\mu\nu})\>+\gamma_{K,S^\dagger S}|\vev{F_S}|^2\right],
}[MassApproxUV]
where $E = i\sqrt2 g\Tra(Q^\dagger\lambda\psi -
\tilde{\psi}\lambda\tQ^\dagger)$.  Note that $m_{\text{sfermion}}^2$ is of
course real, although this is not manifest in \MassApproxUV, a consequence
of the fact that $\cQb^2\cQ^2(K)$ is not manifestly real.

Finally, for a strongly-coupled theory it is more natural to express the
approximations \MassApproxUV in terms of vacuum condensates of IR
elementary fields, i.e.\ the MSSM-restricted mesonic superfield
$\cM=\Tra(M)_{(0,N)}$ and the ``glueball'' superfield
$\cG=-(g^2/32\pi^2)\tra W^\alpha W_\alpha$, leading to
\eqna{
M_{\text{gaugino}} & \approx\frac{\pi\alpha_\text{SM}}{2N|\xi\vev{S}|^2}
\left[2\xi^*\gamma_{K,K}\<S^\dagger\cM^\dagger\>-
2N\gamma_{K,K}\<\cG^\dagger\>-\gamma_{K,S^\dagger S}
\<S^\dagger F_S\>\right],\\
m_{\text{sfermion}}^2 & \approx\frac{C_2(R)\alpha_\text{SM}^2}
{4N|\xi\vev{S}|^2}\left[-2\xi^*\gamma_{K,K}\<S^\dagger
F^\dagger_{\cM}+F^\dagger_S\cM^\dagger\>+N\gamma_{K,K}
\<F^{\vphantom{\dagger}}_{\cG}+F_{\cG}^\dagger\>+\gamma_{K,S^\dagger S}
\vev{F_S^\dagger F_S}\right].
}[MassApproxIR]

Equations \MassApprox, \MassApproxUV and \MassApproxIR can be easily
generalized to more complicated UV theories with several gauge groups and
matter fields in different representations.  They can also be generalized
to closely-related types of mediation like general gaugino
mediation~\cite{Sudano:2010vt}.  The approximations \MassApproxIR are
especially useful since they give an estimate for the MSSM soft
SUSY-breaking masses from the knowledge of the vevs of a few IR elementary
fields, taking the anomalous-dimension matrix to be of $\mathcal{O}(1)$.
Indeed only the vacuum structure of both the messenger and the hidden
sector is necessary to approximately determine the MSSM superpartner
spectrum.  The knowledge of the spectrum of messengers does not directly
enter the computation.

Note that these approximations should be valid for strongly-coupled
theories as well, although the size of the error introduced by truncating
the OPE and assuming that cuts extend to the supersymmetric threshold is
difficult to estimate in general. The anomalous-dimension terms cannot be
computed at strong coupling, but they are expected to be $\mathcal O(1)$.
It is however possible to argue for the functional dependence of the
relevant anomalous dimensions at strong coupling.  For example,
$\gamma_{K,K}$ should depend on the electric quark mass and electric
strong-coupling scale, and since it must be dimensionless it should be
expressible by a series in positive powers of
$|\xi\vev{S}/\Lambda_\text{e}|$ and $|\xi\vev{F_S}/\Lambda_\text{e}^2|$.
For $|\vev{F_S}/\xi\vev{S}|\ll1$, at lowest order one thus
expects\foot{Note that the form of the anomalous current $K$ is known in
terms of magnetic variables around the free supersymmetric and R-symmetric
IR CFT in massless SQCD as described by Seiberg duality~\cite{Abel:2011wv}
(see also \cite{Luty:1999qc}).  However the anomalous dimension computed
from this perspective does not lead to the appropriate functional
dependence as argued here since we are interested in the ISS SUSY-breaking
vacuum.}
\eqn{\gamma_{K,K}\xrightarrow[\text{coupling}]{\text{strong}}\frac{\tNc}{16\pi^2}\left|\frac{\xi\vev{S}}{\Lambda_\text{e}}\right|\delta_{K,K},\qquad\gamma_{K,S^\dagger S}\xrightarrow[\text{coupling}]{\text{strong}}\frac{N\tNc}{16\pi^2}|\xi^2|\delta_{K,S^\dagger S},}[AnomMatStrong]
where $\delta_{K,K}$ and $\delta_{K,S^\dagger S}$ are dimensionless numbers
of order one.  We introduced in \AnomMatStrong one-loop factors as well as
factors of $\tNc$ and $N$ to account for the effective number of degrees of
freedom propagating in the loops as suggested by the Seiberg dual (see
\eqref{Mag}).

Furthermore, although the vevs of the appropriate fields in the vacuum of
interest are not always calculable in the strongly-coupled regime, it is
often possible to approximate them in terms of the relevant scales of the
theory under consideration. Therefore the approximations \MassApproxIR,
which represent the main results of this paper, as well as their
generalizations to more complicated models, should be acceptable up to
dimensionless numbers of order one. Finally, when weakly-coupled duals
exist, it is possible to assess the issues discussed above and directly
check that the approximations \MassApproxIR are indeed reliable up to
$\mathcal{O}(1)$ factors, as will be seen in the next section.

In the event that SUSY is discovered at the LHC and that gauge mediation is
the relevant means of SUSY-breaking communication, the approximations
\MassApproxIR open a rare window into the messenger and the hidden sector:
by experimentally measuring the MSSM superpartner spectrum, they allow an
approximate determination of some of the vevs of operators in the messenger
and the hidden sector.  This is reminiscent of QCD sum
rules~\cite{Shifman:1978bx} (see also~\cite{Colangelo:2000dp} for a nice
review and more references), although here the spectrum of hidden-sector
resonances is not necessary.

We will now use these equations to investigate the superpartner spectra of
sSQCD and its different limits, starting from the computationally-reachable
weakly-coupled regime and ending with the often incalculable
strongly-coupled regime.  To this end, we will use Seiberg
duality~\cite{Seiberg:1994pq}, which for $SU(N_c)$ sSQCD in the free
magnetic phase leads to the following $SU(\tNc\equiv N_f-N_c)$ weakly-coupled
dual theory for the matter fields (here the meson $M$, the magnetic quarks $q$
and $\tilde{q}$, and the singlet $S$),
\eqna{K_\text{m}&=\frac{1}{\alpha|\Lambda_\text{e}|^2}
\Tra(M^\dagger M)_{(0,N_f)}+\frac{1}{\beta}
\Tra(q^\dagger q+\tilde{q}\tilde{q}^\dagger)_{(N_f-N_c,N_f)}+S^\dagger S+\cdots,\\
W_\text{m}&=\frac{1}{\Lambda_\text{d}}\Tra(qM\tilde{q})_{(N_f-N_c,
N_f)}+\xi S\Tra(M)_{(0,N_f)},\\
(-1)^{N_f-N_c}\Lambda_\text{d}^{N_f}&=\Lambda_\text{e}^{3N_c-N_f}
\Lambda_\text{m}^{3(N_f-N_c)-N_f}.}[Mag]
Note that $\alpha$ and $\beta$ are positive real dimensionless numbers of
order one, and $\Lambda_\text{e}$, $\Lambda_\text{m}$ and
$\Lambda_\text{d}$ are the electric strong-coupling scale, the magnetic
scale and the duality scale respectively.\foot{Due to the freedom in
defining the magnetic quarks, $\beta$, $\Lambda_\text{m}$ and $\Lambda_\text{d}$
are not fully determined by the electric theory.} Seiberg duality will allow the
determination of the vevs of the relevant IR elementary fields in terms of a few
unknowns, therefore providing a direct check of the approximations \MassApproxIR.

\subsec{sSQCD in the \texorpdfstring{$g\to0$}{g->0} limit: MGM}[MGM]
In the limit of vanishing hidden-sector gauge coupling, sSQCD is equivalent
to MGM with $N_c$ messenger flavors.  In this limit the phenomenology of
sSQCD is already well-known, and is easily reproduced with our methods.
Indeed, the only non-vanishing vacuum condensate occurs for the MGM singlet
$S$ and the theory is effectively equivalent to MGM with $N_c$ flavors of
messengers as expected.  The approximations \MassApproxIR along with the
one-loop anomalous-dimension matrix \AnomMatWeak thus give (here
$x_S=|\vev{F_S}/\xi\vev{S}^2|$)
\eqna{
M_{\text{gaugino}} &
\approx-\frac{\alpha_\text{SM}}{4\pi}\frac{\vev{F_S}}{\vev{S}}\times
N_c\times\left\{g_\text{approx}(x_S)=\frac{1}{2}\right\},\\
m_{\text{sfermion}}^2 & \approx2\left(\frac{\alpha_\text{SM}}{4\pi}\right)^2
\left|\frac{\vev{F_S}}{\vev{S}}\right|^2\times C_2(R)\times N_c\times\left\{
f_\text{approx}(x_S)=\frac{1}{2}\right\},
}
which, as already mentioned, are only a factor of 2 smaller than the
usually quoted one- and two-loop answers in the limit where $x_S=0$~\cite{Martin:1996zb},
\eqna{g(x_S)&=\frac{1+x_S}{x_S^2}\ln(1+x_S)+\{x_S\to-x_S\}=1+\frac{x_S^2}{6}+\cdots,\\
f(x_S)&=\frac{1+x_S}{x_S^2}\left[\ln(1+x_S)-2\LiTwo\left(\frac{x_S}{1+x_S}\right)+\frac{1}{2}\LiTwo\left(\frac{2x_S}{1+x_S}\right)\right]+\{x_S\to-x_S\}=1+\frac{x_S^2}{36}+\cdots,}
where $\LiTwo(x)=-\int_0^1\,dt\,\frac{\ln(1-xt)}{t}$ is the dilogarithm or
Spence function.  Note that since the OPE is truncated at lowest order in
the SUSY-breaking expansion, it is naturally expected that the approximations
\MassApproxIR only capture (part of) the $x_S=0$ limit of $g(x_S)$ and
$f(x_S)$.

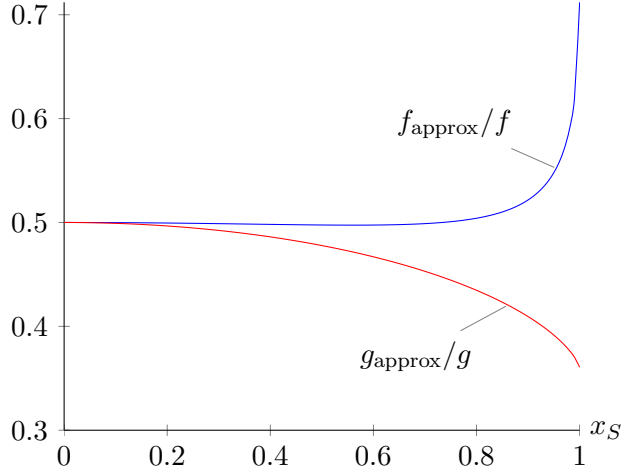
\begin{figure}[ht]
  \centering
  \begin{tikzpicture}
    \begin{axis}[standard,xlabel=$x_S$,ymin=0.3,xmin=0,axis line
      style={-}]
      \addplot+[smooth,mark=none] coordinates {
        (0.,0.5)  (0.01,0.499999)  (0.02,0.499994)  (0.03,0.499988)
        (0.04,0.499978)  (0.05,0.499965)  (0.06,0.49995)  (0.07,0.499932)
        (0.08,0.499912)  (0.09,0.499888)  (0.1,0.499862)  (0.11,0.499834)
        (0.12,0.499803)  (0.13,0.499769)  (0.14,0.499733)  (0.15,0.499694)
        (0.16,0.499653)  (0.17,0.499609)  (0.18,0.499564)  (0.19,0.499516)
        (0.2,0.499465)  (0.21,0.499413)  (0.22,0.499359)  (0.23,0.499303)
        (0.24,0.499244)  (0.25,0.499185)  (0.26,0.499123)  (0.27,0.49906)
        (0.28,0.498995)  (0.29,0.498929)  (0.3,0.498862)  (0.31,0.498794)
        (0.32,0.498725)  (0.33,0.498655)  (0.34,0.498585)  (0.35,0.498514)
        (0.36,0.498443)  (0.37,0.498373)  (0.38,0.498302)  (0.39,0.498232)
        (0.4,0.498162)  (0.41,0.498093)  (0.42,0.498026)  (0.43,0.49796)
        (0.44,0.497897)  (0.45,0.497835)  (0.46,0.497776)  (0.47,0.49772)
        (0.48,0.497667)  (0.49,0.497619)  (0.5,0.497574)  (0.51,0.497535)
        (0.52,0.497501)  (0.53,0.497474)  (0.54,0.497453)  (0.55,0.49744)
        (0.56,0.497435)  (0.57,0.497439)  (0.58,0.497453)  (0.59,0.497479)
        (0.6,0.497517)  (0.61,0.497568)  (0.62,0.497634)  (0.63,0.497716)
        (0.64,0.497816)  (0.65,0.497936)  (0.66,0.498076)  (0.67,0.498241)
        (0.68,0.498431)  (0.69,0.49865)  (0.7,0.4989)  (0.71,0.499185)
        (0.72,0.499508)  (0.73,0.499873)  (0.74,0.500285)  (0.75,0.500749)
        (0.76,0.501271)  (0.77,0.501858)  (0.78,0.502516)  (0.79,0.503256)
        (0.8,0.504087)  (0.81,0.505021)  (0.82,0.506073)  (0.83,0.507259)
        (0.84,0.508599)  (0.85,0.510116)  (0.86,0.511841)  (0.87,0.513808)
        (0.88,0.516062)  (0.89,0.51866)  (0.9,0.521675)  (0.91,0.525203)
        (0.92,0.529374)  (0.93,0.534367)  (0.94,0.540444)  (0.95,0.548001)
        (0.96,0.557682)  (0.97,0.570628)  (0.98,0.589195)  (0.99,0.619712)
        (1.,0.711981)} node
        [pos=0.85,pin={150:{\color{black}{$f_\text{approx}/f$}}},inner
        sep=0pt] {};
      \addplot+[smooth,mark=none] coordinates {
        (0.,0.5)  (0.01,0.499992)  (0.02,0.499967)  (0.03,0.499925)
        (0.04,0.499867)  (0.05,0.499792)  (0.06,0.4997)  (0.07,0.499591)
        (0.08,0.499466)  (0.09,0.499324)  (0.1,0.499165)  (0.11,0.498989)
        (0.12,0.498796)  (0.13,0.498586)  (0.14,0.498359)  (0.15,0.498115)
        (0.16,0.497854)  (0.17,0.497575)  (0.18,0.497279)  (0.19,0.496966)
        (0.2,0.496635)  (0.21,0.496286)  (0.22,0.49592)  (0.23,0.495536)
        (0.24,0.495134)  (0.25,0.494713)  (0.26,0.494275)  (0.27,0.493818)
        (0.28,0.493343)  (0.29,0.492848)  (0.3,0.492336)  (0.31,0.491804)
        (0.32,0.491252)  (0.33,0.490682)  (0.34,0.490092)  (0.35,0.489482)
        (0.36,0.488852)  (0.37,0.488202)  (0.38,0.487531)  (0.39,0.48684)
        (0.4,0.486128)  (0.41,0.485394)  (0.42,0.484639)  (0.43,0.483863)
        (0.44,0.483064)  (0.45,0.482242)  (0.46,0.481398)  (0.47,0.48053)
        (0.48,0.479639)  (0.49,0.478724)  (0.5,0.477785)  (0.51,0.476821)
        (0.52,0.475831)  (0.53,0.474816)  (0.54,0.473774)  (0.55,0.472705)
        (0.56,0.47161)  (0.57,0.470486)  (0.58,0.469333)  (0.59,0.468151)
        (0.6,0.466939)  (0.61,0.465696)  (0.62,0.464421)  (0.63,0.463114)
        (0.64,0.461774)  (0.65,0.460399)  (0.66,0.458988)  (0.67,0.457541)
        (0.68,0.456057)  (0.69,0.454533)  (0.7,0.452969)  (0.71,0.451363)
        (0.72,0.449713)  (0.73,0.448018)  (0.74,0.446277)  (0.75,0.444486)
        (0.76,0.442643)  (0.77,0.440747)  (0.78,0.438794)  (0.79,0.436782)
        (0.8,0.434707)  (0.81,0.432565)  (0.82,0.430353)  (0.83,0.428066)
        (0.84,0.425698)  (0.85,0.423245)  (0.86,0.420697)  (0.87,0.418049)
        (0.88,0.415291)  (0.89,0.41241)  (0.9,0.409395)  (0.91,0.406229)
        (0.92,0.402892)  (0.93,0.399357)  (0.94,0.395591)  (0.95,0.391547)
        (0.96,0.387157)  (0.97,0.382315)  (0.98,0.376837)  (0.99,0.370314)
        (1.,0.360674)} node
        [pos=0.85,pin={220:{\color{black}{$g_\text{approx}/g\!$}}},inner
        sep=0pt] {};
    \end{axis}
  \end{tikzpicture}
  \caption{$g_\text{approx}/g$ and $f_\text{approx}/f$
  as functions of $x_S$ for MGM.}\label{fig:MGMgf}
\end{figure}

The functions $g(x_S)$ and $f(x_S)$, which are only defined in the region
$0\leq x_S\leq 1$ in order to avoid tachyonic messengers, do not deviate much from
unity, and so the agreement of the OPE with the full answer at one loop for the
gauginos and at two loops for the sfermions is reasonable, as can be seen in
Fig.~\ref{fig:MGMgf}.

A complete OPE analysis of MGM shows that the method described
in~\cite{Fortin:2011ad} and extended here works in the weakly-coupled
regime, providing a useful consistency check. Note that it is not easy to
use our method to obtain exact results in the weakly-coupled regime.
Nevertheless, the simple approximations \MassApproxIR match weakly-coupled
computations up to dimensionless numbers of order one, a property which
should translate to the strongly-coupled regime as well.

\subsec{sSQCD in the \texorpdfstring{$\vev{S}\to m/\xi$}{S->m/xi} and
\texorpdfstring{$\vev{F_S}\to0$}{FS->0} limit: mSQCD}
In the limit where the MGM singlet $S$ is assumed frozen without an F-term,
sSQCD is nothing else than mSQCD.  The theory is most interesting in the
free magnetic phase, given by $N_c+1\leq N_f<3N_c/2$, where both a
SUSY-preserving phase and a (metastable) SUSY-breaking phase exist~\cite{Intriligator:2006dd}.

\subsubsec{Around the SUSY vacuum}
In mSQCD, although $\<\cM\>$ and $\<\cG\>$ do not vanish at the
supersymmetric vacuum, the soft SUSY-breaking masses vanish, as expected,
due to the Konishi anomaly~\cite{Konishi:1983hf}.  Indeed, although
\eqn{\frac{g^2}{32\pi^2}\<\tra\lambda\lambda\>=
[\Lambda_\text{e}^{3N_c-N_f}\det(\xi\vev{S})]^\frac{1}{N_c}e^{2\pi ik/N_c},}[vevGlue]
where \vevGlue is valid for any $N_c$ and $N_f$~\cite{Amati:1984uz}, the
vacuum condensate for the mesonic superfield is
\eqn{\<\Tra(\tQ_{\tilde{\imath}}Q^i)_{(N_c,0)}\>=
[\Lambda_\text{e}^{3N_c-N_f}\det(\xi\vev{S})]^\frac{1}{N_c}\left[(\xi\vev{S})^{-1}\right]_
{\tilde{\imath}}^ie^{2\pi ik/N_c},}[vevMeson]
as enforced by the Konishi anomaly~\cite{Konishi:1983hf},
\eqn{\frac{-i}{2\sqrt{2}}\{\bar{\cQ}_{\dot{\alpha}},
\Tra(\bar{\psi}_{\tilde{\imath}}^{\dot{\alpha}}Q^i)_{(N_c,0)}\}=-
\xi\vev{S}\Tra(\tQ_{\tilde{\imath}}Q^i)_{(N_c,0)}+
\delta_{\tilde{\imath}}^i\frac{g^2}{32\pi^2}\tra\lambda\lambda,}[Konishi]
in supersymmetric vacua.\foot{Here the index $k$ labels the degenerate SUSY
vacua which arise from the spontaneous breaking of the discrete global
symmetry $\mathbb{Z}_{2N_c}$ to $\mathbb{Z}_2$.}  In terms of the IR fields
this implies that $\xi\vev{S}\<\cM\>=N\<\cG\>$.  Since all remaining vacuum
condensates vanish, the approximations \MassApproxIR lead to a superpartner
spectrum consistent with SUSY.

\begin{figure}[ht]
  \centering
  \begin{tikzpicture}[>=latex]
    \draw[->,line cap=round,line join=round] (0,0)--(7.5,0) node[below] {$\Phi$};
    \draw[->,line cap=round,line join=round] (0,0)--(0,3.5) node[left=-1pt] {$V$};
    \draw[line cap=round,line join=round] plot[smooth] coordinates {(0,1)
    (0.35,1.2) (1,2.7) (1.8,3) (2.8,2.4) (5.7,0.4pt) (6.8,1.5)};
  \end{tikzpicture}
  \caption{A sketch of the potential of mSQCD.} \label{fig:mSQCDPotential}
\end{figure}
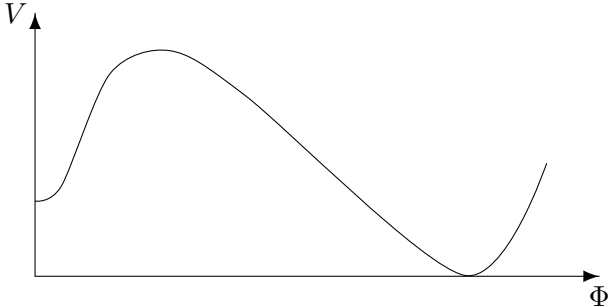

\subsubsec{Around the ISS vacuum}
As shown by ISS~\cite{Intriligator:2006dd}, mSQCD with small masses has a
metastable SUSY-breaking minimum close to the origin of field space.  A sketch
of the potential of mSQCD is shown in Fig.~\ref{fig:mSQCDPotential}.

Since the SUSY-breaking scale and the messenger scale are the same in ISS,
there is no dimensionless SUSY-breaking parameter to keep track of the order
at which SUSY-breaking effects appear in any computation.  Thus, in order to
compare \MassApproxIR with weakly-coupled computations of the sfermion masses
from the dual theory (see Appendix~\ref{app:Supspec}), it is convenient to
distinguish between the SUSY-breaking scale and the messenger scale by
introducing two $\xi$'s, $(\xi,\xi_L)$ with $x_\cM=\xi_L/\xi$ and
$0\leq|x_\cM|\leq1$. This effectively splits the mass matrix in two sectors
and allows us to keep track of the SUSY-breaking effects.

The location of the SUSY-breaking minimum can be found using the dual theory \Mag
and, in terms of the IR elementary fields (embedding the MSSM into the $X$-sector
of \eqref{ansatz}), is given by
\eqn{\vev{\mathcal M}=\vev{\mathcal{G}}=\vev{F_\mathcal{G}}=0,\qquad\vev{F_\mathcal M}=-N_c\alpha\xi_L^*\vev{S^\dagger}|\Lambda_\text{e}|^2.}[vevISS]
The ISS vacuum faces an immediate problem for phenomenological
applications: it has an accidental R-symmetry and thus constrains to zero
Majorana gaugino masses.  This can be seen directly from the approximations
\MassApproxIR and the vevs \vevISS.  The sfermion masses, on the other
hand, are not constrained by the accidental R-symmetry and are indeed
non-zero, as is also clear from \MassApproxIR and the vevs \vevISS.

\begin{figure}[ht]
  \centering
  \begin{tikzpicture}
    \begin{axis}[standard,xlabel=$|x_\mathcal{M}|$,ylabel=$f_{\text{approx}}/f$,
        ymin=0.24,xmin=0,axis line style={-}]
      \addplot+[smooth,mark=none] coordinates {
        (0.,0.25)  (0.01,0.249999)  (0.02,0.249997)  (0.03,0.249994)
        (0.04,0.249989)  (0.05,0.249983)  (0.06,0.249975)  (0.07,0.249966)
        (0.08,0.249956)  (0.09,0.249944)  (0.1,0.249931)  (0.11,0.249917)
        (0.12,0.249901)  (0.13,0.249884)  (0.14,0.249866)  (0.15,0.249847)
        (0.16,0.249826)  (0.17,0.249805)  (0.18,0.249782)  (0.19,0.249758)
        (0.2,0.249733)  (0.21,0.249707)  (0.22,0.249679)  (0.23,0.249651)
        (0.24,0.249622)  (0.25,0.249592)  (0.26,0.249561)  (0.27,0.24953)
        (0.28,0.249498)  (0.29,0.249465)  (0.3,0.249431)  (0.31,0.249397)
        (0.32,0.249363)  (0.33,0.249328)  (0.34,0.249293)  (0.35,0.249257)
        (0.36,0.249222)  (0.37,0.249186)  (0.38,0.249151)  (0.39,0.249116)
        (0.4,0.249081)  (0.41,0.249047)  (0.42,0.249013)  (0.43,0.24898)
        (0.44,0.248948)  (0.45,0.248917)  (0.46,0.248888)  (0.47,0.24886)
        (0.48,0.248834)  (0.49,0.248809)  (0.5,0.248787)  (0.51,0.248768)
        (0.52,0.248751)  (0.53,0.248737)  (0.54,0.248726)  (0.55,0.24872)
        (0.56,0.248717)  (0.57,0.248719)  (0.58,0.248727)  (0.59,0.248739)
        (0.6,0.248758)  (0.61,0.248784)  (0.62,0.248817)  (0.63,0.248858)
        (0.64,0.248908)  (0.65,0.248968)  (0.66,0.249038)  (0.67,0.24912)
        (0.68,0.249216)  (0.69,0.249325)  (0.7,0.24945)  (0.71,0.249593)
        (0.72,0.249754)  (0.73,0.249937)  (0.74,0.250143)  (0.75,0.250375)
        (0.76,0.250636)  (0.77,0.250929)  (0.78,0.251258)  (0.79,0.251628)
        (0.8,0.252044)  (0.81,0.252511)  (0.82,0.253037)  (0.83,0.25363)
        (0.84,0.254299)  (0.85,0.255058)  (0.86,0.25592)  (0.87,0.256904)
        (0.88,0.258031)  (0.89,0.25933)  (0.9,0.260838)  (0.91,0.262602)
        (0.92,0.264687)  (0.93,0.267183)  (0.94,0.270222)  (0.95,0.274001)
        (0.96,0.278841)  (0.97,0.285314)  (0.98,0.294597)  (0.99,0.309856)
        (1,0.35599)};
    \end{axis}
  \end{tikzpicture}
  \caption{$f_\text{approx}/f$ as function of $|x_\cM|$ for
  mSQCD with $\beta=\delta_{K,K}=1$.  Both $g_\text{approx}(x_\cM)$ and
  $g(x_\cM)$ vanish and so the corresponding ratio is not plotted here.}\label{fig:mSQCDgf}
\end{figure}
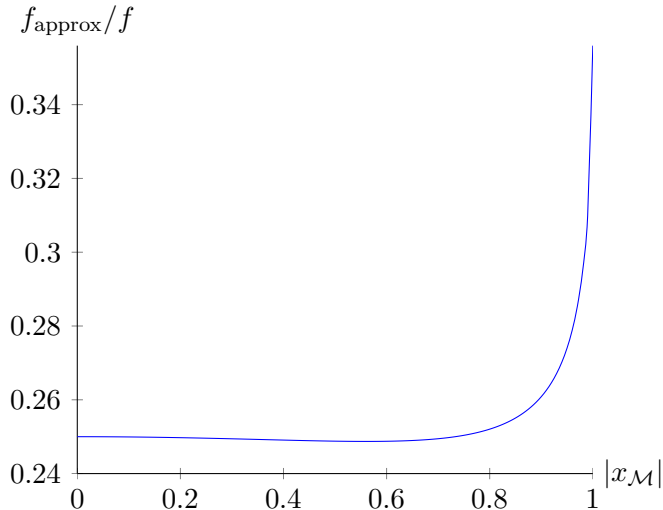

Fixing $\Lambda_\text{d}=\Lambda_\text{m}=(-1)^{(N_c-N_f)/(3N_c-N_f)}\Lambda_\text{e}$
and using the anomalous dimensions \AnomMatStrong the approximated sfermion
masses obtained from \MassApproxIR are
\eqn{m_{\text{sfermion}}^2\approx2\left(\frac{\alpha_\text{SM}}{4\pi}\right)^2|x_\cM|^2\alpha\beta|\xi\vev{S}\Lambda_\text{e}|\times C_2(R)\times\tNc\times\left\{f_\text{approx}(x_\cM)=\frac{4\pi^2}{\tNc\beta}
\left|\frac{\Lambda_\text{e}}{\xi\vev{S}}\right|\gamma_{K,K}=\frac{\delta_{K,K}}{4\beta}\right\},}
while using the dual theory the weakly-coupled computation gives
\eqna{f(x_\cM) &=\frac{1+|x_\cM|}{|x_\cM|^2}\left[\ln(1+|x_\cM|)-2\LiTwo\left(\frac{|x_\cM|}{1+|x_\cM|}\right)+\frac{1}{2}\LiTwo\left(\frac{2|x_\cM|}{1+|x_\cM|}\right)\right]+\{|x_\cM|\to-|x_\cM|\}\\
&=1+\frac{|x_\cM|^2}{36}+\cdots.}
Although $x_\cM=1$ in ISS, our approximations only rely on the
lowest-order operators appearing in the OPE and should only capture (part of)
the $x_\cM=0$ contributions to $f(x_\cM)$, up to a number of order one
(as in the MGM case of section~\ref{MGM}).  This is exactly what happens
here.  Moreover, since the function $f(x_\cM)$ stays close to unity for all
$x_\cM$, the approximations \MassApproxIR are reasonable for $0\leq|x_\cM|\leq1$
as shown in Fig.~\ref{fig:mSQCDgf}.
Therefore the method developed here gives sensible results even in strongly-coupled
theories including higher-order SUSY-breaking corrections.

It is interesting to notice that a full knowledge of the OPE could
possibly lead to a computation of the anomalous dimensions of relevant
operators in mSQCD following the method described here, as was done for MGM
in~\cite{Fortin:2011ad}.  For more details the reader is referred to
section~\ref{sSQCDISS} and~\cite{Intriligator:2006dd}.

\subsec{sSQCD in the free magnetic phase}
Here we explore sSQCD for $N_c+1\leq N_f<3N_c/2$.  As mentioned above, in
mSQCD dynamical SUSY breaking in metastable vacua occurs for this range of
$N_f$ close to the origin of field space~\cite{Intriligator:2006dd}.
Although the electric theory is strongly coupled, Seiberg duality allows
one to establish the presence of SUSY breaking.  In this subsection we also
use Seiberg duality to understand SUSY breaking in sSQCD close to the
origin of field space.

\subsubsec{Around the would-be SUSY vacuum}
Let us first discuss the fate of the would-be SUSY vacuum of mSQCD in the
full sSQCD theory.  For $|\vev{F_S}/\xi\vev{S}^2|\ll1$ one would expect that
the vevs of the glueball and mesonic superfields are only slightly perturbed
compared to their mSQCD values \vevGlue and \vevMeson.  Moreover, from the
point of view of the sSQCD fields, SUSY is explicitly broken.  One should
thus expect that the SUSY vacuum of mSQCD becomes a SUSY-breaking vacuum in
sSQCD.  Since small instantons are relevant, it is impossible to compute
the vevs of the glueball and mesonic fields from instanton techniques
without a full knowledge of the hidden-sector theory.  It is nevertheless
possible to estimate the vev of the lowest component of the mesonic
superfield from the superpotential and the K\"{a}hler potential \Mag,
leading to
\eqn{\<\cM\>=[\xi^{N_f-N_c}\vev{S}^{N_f-N_c}
\Lambda_\text{e}^{3N_c-N_f}]^\frac{1}{N_c}
\left[1+\frac{N_c-N_f}{N_c}\frac{1}{\alpha|\xi|^2}
\left(\frac{\xi^*\vev{S^\dagger}}{\Lambda^*_\text{e}}\right)
^{\frac{N_f}{N_c}}\frac{\vev{F_S^\dagger}\Lambda_\text{e}^{*2}}
{\vev{S^{\dagger}}^2\vev{S}\Lambda_\text{e}}+\cdots\right].}
One could then use the Konishi anomaly \Konishi to obtain the vev of the
glueball superfield, but since the vacuum is expected to be
non-supersymmetric, this approach is inconclusive.  A complete knowledge of
the hidden sector seems thus necessary to determine the characteristics of
the superpartner spectrum around this vacuum.

\subsubsec{Around the ISS-like vacuum}[sSQCDISS]
Around the origin of field space it is more convenient to use the dual theory
as given by \Mag, but with canonically-normalized matter fields $\Phi$, $\varphi$
and $\tvarphi$.  The superpotential becomes
\eqn{W_\text{m}=h\Tra \varphi\Phi\tvarphi-h\Psi\Tra\Phi,}
where $\Psi$ is a background field with $\vev{\Psi}=\mu^2+\theta^2\mu_F^3$.
The parameter $\mu_F$ is the source of R-symmetry breaking in our example.
With the parametrization
\eqn{\Phi=\begin{pmatrix}
  Y_{\tNc\times\tNc} & Z^T_{\tNc\times N_c}\\
  \tZ_{N_c\times\tNc} & X_{N_c\times N_c}
\end{pmatrix},\qquad
\varphi^T=\begin{pmatrix}
  \chi_{\tNc\times\tNc}\\
  \rho_{N_c\times\tNc}
\end{pmatrix},\qquad
\tvarphi=\begin{pmatrix}
  \tchi_{\tNc\times\tNc}\\
  \trho_{N_c\times\tNc}
\end{pmatrix},}[ansatz]
the scalar potential becomes
\eqna{V&=N_f|h\mu^2|^2+h\mu_F^3\Tra(Y+X)+h^*\mu_F^{*3}
\Tra(Y^\dagger+X^\dagger)\\
&\quad+|h|^2\Tra[-\mu^2(\tchi^\dagger\chi^*+\trho^\dagger\rho^*)
-\mu^{* 2}(\chi^T\tchi+\rho^T\trho)\\
&\hspace{1.95cm}+\tchi^\dagger(Y^\dagger Y+\tZ^\dagger \tZ)\tchi
+\trho^\dagger(Z^*Z^T+X^\dagger X)\trho
+\trho^\dagger(Z^*Y+X^\dagger\tZ)\tchi\\
&\hspace{1.95cm}+\tchi^\dagger(Y^\dagger Z^T+\tZ^\dagger X)\trho
+\chi^\dagger(Y^*Y^T+Z^\dagger Z)\chi
+\rho^\dagger(\tZ^*\tZ^T+X^*X^T)\rho\\
&\hspace{1.95cm}+\rho^\dagger(\tZ^*Y^T+X^*Z)\chi
+\chi^\dagger(Y^*\tZ^T+Z^\dagger X^T)\rho
+(\chi^T\chi^*+\rho^T\rho^*)(\tchi^\dagger\tchi+\trho^\dagger\trho)].}
As in the ISS case, the rank condition implies that SUSY is broken with
$F_X^\dagger=h\mu^2$, and a minimum should develop around the origin of
field space, which can be conveniently described with the following ansatz:
\eqn{\vev{\Phi}=\begin{pmatrix}
  Y_0 & 0\\
  0 & X_0
\end{pmatrix},\qquad
\vev{\varphi^T}=\begin{pmatrix}
  q_0\\
  0
\end{pmatrix},\qquad
\vev{\tvarphi}=\begin{pmatrix}
  \tq_0\\
  0
\end{pmatrix}\!.}
Assuming $\tq_0=q_0=q$ the scalar potential is minimized in (almost) all
directions when
\eqna{Y_0&=-\frac{\mu_F^{*3}}{h(|q_0|^2+|\tq_0|^2)}=-\frac{\mu_F^{*3}}{2h|q|^2},\\
q&=\frac13\mu(1+H^{1/3}+H^{-1/3})^{1/2},\qquad\text{where}\qquad
H=1-\frac{27}{2}|\epsilon|\left(|\epsilon|-\sqrt{|\epsilon|^2-\frac{4}{27}}\right).}[vevMag]
Here $\epsilon=\mu_F^3/2h^*\mu^{*2}\mu$ and it is assumed small.  The constraint
on $q$ comes from minimization in the $\tchi$-direction leading to the
condition
\eqn{|q|^6-\mu^{*2}q^2|q|^2+\left|\frac{\mu_F^3}{2h}\right|^2=0,}
which requires that $q/\mu\in\mathbb{R}$.  Keeping the solution\foot{The
other solutions lead to tachyons.} for which $q\xrightarrow[\mu_F\to
0]{}\mu$ leads to the vev mentioned above.  For a well-defined $q$ one
needs $|\epsilon|\leq\frac{2\sqrt{3}}{9}$ which is easily satisfied for
small $|\epsilon|$.  For small $\mu_F$ (or $\epsilon$), \vevMag can be
approximated by
\eqn{Y_0=-\frac{\mu_F^{*3}}{2h|\mu|^2}+\cdots,\qquad
q=\mu\left(1-\frac{1}{2}|\epsilon|^2+\cdots\right).}

The scalar potential is stabilized in all but the $X$-direction.  As opposed
to the ISS case where $X$ is a flat direction of $V$, here $X$ is a runaway
direction at tree level and $V$ slopes down in the $X$-direction.  Since the
runaway behavior is dictated by the small deformation $\mu_F$, it is expected that
the one-loop Coleman--Weinberg potential stabilizes the runaway direction close to
the origin of field space, thus leading to spontaneous breaking of the accidental
R-symmetry of the ISS model and allowing for non-vanishing gaugino masses.

To calculate the Coleman--Weinberg potential~\cite{Coleman:1973jx} for a
general supersymmetric theory with $n$ chiral superfields $\Phi^i$, canonical
K\"{a}hler potential, and superpotential $W(\Phi)$, we need the mass matrices
for scalar and spin-$\tfrac{1}{2}$ fields, given respectively by the
$2n\times 2n$ matrices
\eqn{\bbM_0^2 = \begin{pmatrix}
         W^{\dagger ik}W_{kj} & W^{\dagger ijk}W_k\\
         W_{ijk}W^{\dagger k} & W_{ik}W^{\dagger kj}\\
        \end{pmatrix} \qquad \text{and} \qquad
        \bbM^2_{1/2} = \begin{pmatrix}
                                W^{\dagger ik}W_{kj} & 0\\
                                0 & W_{ik}W^{\dagger kj}
                                \end{pmatrix}\!,}
with $W_i\equiv \partial W/\partial\Phi^i$ and similarly for the rest, where
the derivatives are to be evaluated at the vevs computed for the zero
components of the chiral superfields.

In the case of supersymmetric theories, where quadratic divergences cancel
among bosons and fermions,
\eqn{\STr\bbM^2\equiv\Tra\bbM_0^2-\Tra\bbM_{1/2}^2=0,}[SuperTr]
the Coleman--Weinberg potential takes the form
\eqn{V_\text{CW} = \frac{1}{64\pi^2}\STr \bbM^4
\ln\frac{\bbM^2}{\Lambda^2} \equiv
\frac{1}{64\pi^2}\left[\Tra\bbM_0^4
\left(\ln\frac{\bbM_0^2}{4\Lambda^2}+\frac{1}{2}\right) - \Tra\bbM_{1/2}^4
\left(\ln\frac{\bbM_{1/2}^2}{4\Lambda^2}+\frac{1}{2}\right)\right],}
where $\Lambda$ is the cutoff scale and plays no role in the following.
We are therefore interested in $V_\text{CW}$ as a function of the runaway
direction $X$, $V_\text{CW}(X)$.  Due to the supertrace relation \SuperTr,
we only have to consider the mass matrices for the $(\rho,Z)$ sector, since
this is the only sector in which the spectrum is non-supersymmetric at tree
level.

The mass eigenstates for the messenger sectors are fairly complicated.  To
simplify the analysis we choose to compute them at order $\epsilon$, leading
to
\eqna{\tilde{m}_1^2&=|h\mu|^2\frac{\epsilon x+\epsilon^*x^*}{1+|x|^2},\\
\tilde{m}_2^2&=|h\mu|^2\left(1+|x|^2-\frac{\epsilon x+\epsilon^*x^*}{1+|x|^2}\right),\\
\tilde{m}_3^2&=|h\mu|^2\left(\frac{3}{2}+\frac{1}{2}|x|^2-\frac{1}{2}(1+6|x|^2+|x|^4)^{1/2}\right.\\
&\hspace{2cm}\left.+\frac{1+|x|^2-(1+6|x|^2+|x|^4)^{1/2}}{1+6|x|^2+|x|^4-(1+|x|^2)(1+6|x|^2+|x|^4)^{1/2}}(\epsilon x+\epsilon^*x^*)\right),\\
\tilde{m}_4^2&=|h\mu|^2\left(\frac{3}{2}+\frac{1}{2}|x|^2+\frac{1}{2}(1+6|x|^2+|x|^4)^{1/2}\right.\\
&\hspace{2cm}\left.+\frac{1+|x|^2+(1+6|x|^2+|x|^4)^{1/2}}{1+6|x|^2+|x|^4+(1+|x|^2)(1+6|x|^2+|x|^4)^{1/2}}(\epsilon x+\epsilon^*x^*)\right),}[bosonmess]
for the bosonic mass eigenstates and
\eqna{m_1^2&=|h\mu|^2\left(1+\tfrac{1}{2}|x|^2-\tfrac{1}{2}|x|(4+|x|^2)^{1/2}\right),\\
m_2^2&=|h\mu|^2\left(1+\tfrac{1}{2}|x|^2+\tfrac{1}{2}|x|(4+|x|^2)^{1/2}\right),}[fermmess]
for the fermionic mass eigenstates.  Note that to simplify the notation we
introduced $x=X/\mu$.  Moreover, it is important to notice that $\tilde{m}_1$
vanishes exactly once higher-order corrections are introduced since it corresponds
to a Goldstone boson.

\begin{figure}[ht]
  \centering
  \begin{tikzpicture}[>=latex]
    \draw[->,line cap=round,line join=round] (0,0)--(7.5,0) node[below] {$\Phi$};
    \draw[->,line cap=round,line join=round] (0,0)--(0,3.3) node[left=-1pt] {$V$};
    \draw[line width=7pt,color=gray!40,line cap=round] plot[smooth]
    coordinates {(5.1,0.6) (5.7,0.3) (6.2,0.6) (6.8,1.5)};
    \draw[line cap=round,line join=round] plot[smooth] coordinates {(0,1)
    (0.35,0.9) (1.5,2.7) (5.1,0.6)};
    \draw[line cap=round,line join=round] plot[smooth] coordinates
    {(5.1,0.6) (5.7,0.3) (6.2,0.6) (6.8,1.5)};
  \end{tikzpicture}
  \caption{A sketch of the potential of sSQCD. The shading indicates that
  our analysis of the spectrum in the corresponding region, i.e.\ around
  and past the would-be SUSY vacuum, is not conclusive.}
  \label{fig:sSQCDPotential}
\end{figure}
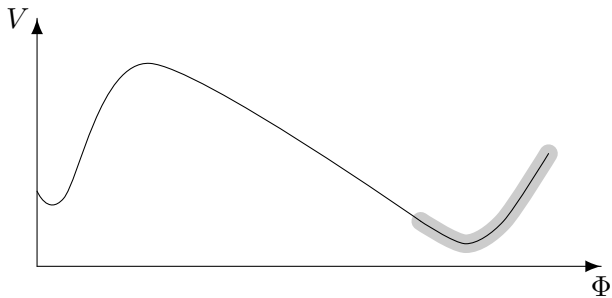

Including the Coleman--Weinberg potential with corrections up to
$\mathcal{O}(\epsilon)$ terms, the runaway in the $X$-direction is found to
be stabilized at
\eqn{X_0=-\frac{16\pi^2+\tNc|h|^2\ln2}{\tNc|h|^2(\ln4-1)}\epsilon^*\mu,}
and a minimum appears close to the origin in field space. As we have
explained, SUSY is also broken in the faraway vacuum. A sketch of the
potential can be seen in Fig.~\ref{fig:sSQCDPotential}.

To make use of \MassApproxIR we relate the canonically-normalized IR fields
to the UV elementary fields with the help of the following dictionary:
\begin{gather*}
\varphi = \frac{q}{\sqrt{\beta}},\qquad
\tvarphi = \frac{\tilde{q}}{\sqrt{\beta}},\qquad
\Phi = \frac{M}{\sqrt{\alpha}\Lambda_\text{e}},\\
h=\frac{\sqrt{\alpha}\beta\Lambda_\text{e}}{\Lambda_\text{d}},\qquad
\mu^2=-\frac{\xi\vev{S}\Lambda_\text{d}}{\beta},\qquad
\mu_F^3=-\frac{\xi\vev{F_S}\Lambda_\text{d}}{\beta}.
\end{gather*}
As already mentioned, one can choose to fix $\Lambda_\text{d}=\Lambda_\text{m}=(-1)^{(N_c-N_f)/(3N_c-N_f)}\Lambda_\text{e}$
and describe the results in terms of $\alpha$ and $\beta$, which leads to ($N=N_c$)
\eqn{\vev{\cM}=\frac{N_c\sigma(x_\cM)}{2\beta}\left|\frac{\Lambda_\text{e}}{\xi\vev{S}}\right|\xi^*\vev{F_S^\dagger},\qquad\vev{F_\cM}=-N_c\alpha\xi_L^*\vev{S^\dagger}|\Lambda_\text{e}|^2,}
when embedding the MSSM gauge group into the $X$-sector of \ansatz.  Here
$\xi_L$ has been introduced, as in the ISS case, to keep track of the
SUSY-breaking effects, and $\sigma(x_\cM)$ encodes the position of the
minimum as a function of the SUSY-breaking effects,
\eqna{\sigma(x_\cM)&=\frac{16\pi^2+\tNc\alpha\beta^2a}{\tNc\alpha\beta^2b}x_\cM^*,\\
a&=\frac{1}{2x_\cM}\big[(1+|x_\cM|)\ln(1+|x_\cM|)+\{|x_\cM|\to-|x_\cM|\}\big]\xrightarrow[x_\cM\to1]{}\ln2,\\
b&=\frac{1}{2|x_\cM|}\big[(1+|x_\cM|)^2\ln(1+|x_\cM|)-|x_\cM|-\{|x_\cM|\to-|x_\cM|\}\big]\xrightarrow[x_\cM\to1]{}\ln4-1.}
Using the anomalous dimensions \AnomMatStrong, the superpartner spectrum at order
$\mathcal{O}(\mu_F^3)\sim\mathcal{O}(\vev{F_S})$ is thus
\eqna{
M_{\text{gaugino}} &
\approx\frac{\alpha_\text{SM}}{4\pi}\frac{\vev{F_S}}{\vev{S}}\times\tNc\times\left\{g_\text{approx}(x_\cM)=\sigma^*(x_\cM)x_\cM^*\frac{2\pi^2}{\tNc\beta}\left|\frac{\Lambda_\text{e}}{\xi\vev{S}}\right|\gamma_{K,K}-\frac{2\pi^2}{N_c\tNc|\xi|^2}\gamma_{K,S^\dagger S}=\right.\\
&\left.\hspace{9.5cm}=\sigma^*(x_\cM)x_\cM^*\frac{\delta_{K,K}}{8\beta}-\frac{\delta_{K,S^\dagger S}}{8}\right\},\\
m_{\text{sfermion}}^2 & \approx2\left(\frac{\alpha_\text{SM}}{4\pi}\right)^2|x_\cM|^2\alpha\beta|\xi\vev{S}\Lambda_\text{e}|\times C_2(R)\times\tNc\\
&\hspace{7cm}\times\left\{f_\text{approx}(x_\cM)=\frac{4\pi^2}{\tNc\beta}
\left|\frac{\Lambda_\text{e}}{\xi\vev{S}}\right|\gamma_{K,K}=\frac{\delta_{K,K}}{4\beta}
\right\},
}[sSQCDapprox]
and can be compared to the weakly-coupled computation which gives
\eqna{
g(x_\cM) &= \left[\frac{1+|x_\cM|}{|x_\cM|^2}\ln(1+|x_\cM|)+\left\{|x_\cM|\to-|x_\cM|\right\}\right]\\
&\hspace{2cm}+\frac{\sigma^*(x_\cM)}{2x_\cM|x_\cM|}\left[3|x_\cM|-(3+4|x_\cM|+|x_\cM|^2)\ln(1+|x_\cM|)-\left\{|x_\cM|\to-|x_\cM|\right\}\right]\\
&=1+\frac{|x_\cM|^2}{6}+\cdots+\sigma^*(x_\cM)\left(\frac{x_\cM^*|x_\cM|^2}{15}+\cdots\right),\\
f(x_\cM) &=\frac{1+|x_\cM|}{|x_\cM|^2}\left[\ln(1+|x_\cM|)-2\LiTwo\left(\frac{|x_\cM|}{1+|x_\cM|}\right)+\frac{1}{2}\LiTwo\left(\frac{2|x_\cM|}{1+|x_\cM|}\right)\right]+\{|x_\cM|\to-|x_\cM|\}\\
&=1+\frac{|x_\cM|^2}{36}+\cdots.}[fgsSQCD]

At order $\mathcal{O}(\vev{F_S})$ the sSQCD sfermion masses are the same as
the mSQCD sfermion masses.  Note that the functional dependence of the
anomalous dimension $\gamma_{K,K}$, necessary for the approximate gaugino
masses to match the weakly-coupled computation, is the same as the one
expected from the sfermion masses.  This strongly suggests that the
functional dependence of $\gamma_{K,K}$ is indeed proportional to
$|\xi\vev{S}/\Lambda_\text{e}|$.

\begin{figure}[ht]
  \centering
    \begin{tikzpicture}
      \begin{axis}[standard,xlabel=$|x_\mathcal{M}|$,xmin=0.5,ymin=0,axis
        line style={-}]
      \addplot+[smooth,mark=none] coordinates {
        (0.5,0.248787)  (0.505,0.248777)  (0.51,0.248768)  (0.515,0.248759)
        (0.52,0.248751)  (0.525,0.248743)  (0.53,0.248737)  (0.535,0.248731)
        (0.54,0.248726)  (0.545,0.248723)  (0.55,0.24872)  (0.555,0.248718)
        (0.56,0.248717)  (0.565,0.248718)  (0.57,0.248719)  (0.575,0.248722)
        (0.58,0.248727)  (0.585,0.248732)  (0.59,0.248739)  (0.595,0.248748)
        (0.6,0.248758)  (0.605,0.24877)  (0.61,0.248784)  (0.615,0.248799)
        (0.62,0.248817)  (0.625,0.248836)  (0.63,0.248858)  (0.635,0.248882)
        (0.64,0.248908)  (0.645,0.248937)  (0.65,0.248968)  (0.655,0.249002)
        (0.66,0.249038)  (0.665,0.249078)  (0.67,0.24912)  (0.675,0.249166)
        (0.68,0.249216)  (0.685,0.249268)  (0.69,0.249325)  (0.695,0.249386)
        (0.7,0.24945)  (0.705,0.249519)  (0.71,0.249593)  (0.715,0.249671)
        (0.72,0.249754)  (0.725,0.249843)  (0.73,0.249937)  (0.735,0.250037)
        (0.74,0.250143)  (0.745,0.250255)  (0.75,0.250375)  (0.755,0.250501)
        (0.76,0.250636)  (0.765,0.250778)  (0.77,0.250929)  (0.775,0.251089)
        (0.78,0.251258)  (0.785,0.251438)  (0.79,0.251628)  (0.795,0.25183)
        (0.8,0.252044)  (0.805,0.25227)  (0.81,0.252511)  (0.815,0.252766)
        (0.82,0.253037)  (0.825,0.253324)  (0.83,0.25363)  (0.835,0.253954)
        (0.84,0.254299)  (0.845,0.254667)  (0.85,0.255058)  (0.855,0.255475)
        (0.86,0.25592)  (0.865,0.256396)  (0.87,0.256904)  (0.875,0.257448)
        (0.88,0.258031)  (0.885,0.258657)  (0.89,0.25933)  (0.895,0.260055)
        (0.9,0.260838)  (0.905,0.261684)  (0.91,0.262602)  (0.915,0.263599)
        (0.92,0.264687)  (0.925,0.265877)  (0.93,0.267183)  (0.935,0.268625)
        (0.94,0.270222)  (0.945,0.272003)  (0.95,0.274001)  (0.955,0.276261)
        (0.96,0.278841)  (0.965,0.281821)  (0.97,0.285314)  (0.975,0.289486)
        (0.98,0.294597)  (0.985,0.301095)  (0.99,0.309856)  (0.995,0.323093)
        (0.997,0.331104)  (0.999,0.343276)  (1.,0.35599)} node
        [pos=0.4,pin={120:{\color{black}{$f_\text{approx}/f$}}},inner
        sep=0pt] {};
      \addplot+[smooth,mark=none] coordinates {
        (0.5,5.18169)  (0.51,4.98559)  (0.52,4.79772)  (0.53,4.61766)
        (0.54,4.44503)  (0.55,4.27945)  (0.56,4.12057)  (0.57,3.96805)
        (0.58,3.82158)  (0.59,3.68085)  (0.6,3.54559)  (0.61,3.41553)
        (0.62,3.29042)  (0.63,3.17)  (0.64,3.05406)  (0.65,2.94238)
        (0.66,2.83475)  (0.67,2.73099)  (0.68,2.6309)  (0.69,2.53431)
        (0.7,2.44106)  (0.71,2.35099)  (0.72,2.26395)  (0.73,2.17979)
        (0.74,2.09838)  (0.75,2.01958)  (0.76,1.94328)  (0.77,1.86935)
        (0.78,1.79768)  (0.79,1.72815)  (0.8,1.66067)  (0.81,1.59512)
        (0.82,1.5314)  (0.83,1.46942)  (0.84,1.40907)  (0.85,1.35027)
        (0.86,1.29291)  (0.87,1.23689)  (0.88,1.1821)  (0.89,1.12845)
        (0.9,1.07582)  (0.91,1.02407)  (0.92,0.97306)  (0.93,0.922618)
        (0.94,0.872526)  (0.95,0.8225)  (0.96,0.772138)  (0.97,0.720811)
        (0.98,0.66739)  (0.99,0.609268)  (1.,0.531292)}
        node [pos=0.7,pin={85:{\color{black}{$g_\text{approx}/g$}}},inner
        sep=0pt] {};
    \end{axis}
  \end{tikzpicture}
  \caption{$g_\text{approx}/g$ and $f_\text{approx}/f$ as functions of
  $|x_\cM|$ for sSQCD with $\beta=\delta_{K,K}=\delta_{K,S^\dagger
  S}=1$ and $\tNc=2$.}\label{fig:sSQCDgf}
\end{figure}
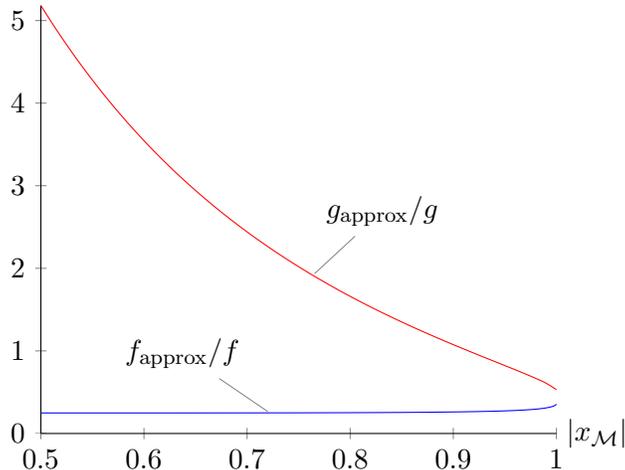

As for the mSQCD case, $x_\cM=1$ but by truncating the OPE the results
\sSQCDapprox should only capture the lowest-order contribution in the
$x_\cM$-expansion of $g(x_\cM)$ and $f(x_\cM)$ up to $\mathcal{O}(1)$
factors, as can be seen directly.  Note however that the power in $|x_\cM|$
of the spontaneous R-symmetry
breaking contribution to the gaugino mass, denoted by $\sigma(x_\cM)$,
does not exactly match the weakly-coupled computation: it is off by a factor of
$|x_\cM|^2$.  This suggests that all OPE contributions of the same type must be
included to appreciate the suppression seen at small dynamical SUSY breaking,
i.e. for small $|x_\cM|$.  Yet, this point is of no relevance since the
metastable SUSY-breaking minimum disappears for small $|x_\cM|$, indeed
$\vev{\cM}\xrightarrow[x_\cM\to0]{}\infty$.  This is clear since for fixed
$\epsilon$, the Coleman--Weinberg potential cannot compete against the runaway
when $|x_\cM|$ is too small.  The value of $x_\cM$ at which the minimum
disappears can be estimated from the constraint that the messenger masses must
be all non-tachyonic.  Using the messenger masses at order $\epsilon$, this
constraint is obtained from the fermionic messenger mass eigenstates \fermmess.
In Fig.~\ref{fig:sSQCDgf} we plot our results for $0.5\leq|x_\mathcal{M}|\leq 1$.

Note that the gaugino approximation overestimates the mass if all
dimensionless numbers are positive.\foot{From the sfermion mass
\sSQCDapprox it is clear that $\delta_{K,K}/\beta$ is positive and thus
$\delta_{K,K}$ must be positive.}  Overall, the method described here
gives sensible results even for strongly-coupled theories of SUSY-breaking.
Again, a complete knowledge of the OPE could allow a determination of the
anomalous dimensions of relevant operators of sSQCD using these methods.

Finally, even though it is not the main purpose of this paper, it is of
interest to discuss some of the phenomenology of this new deformation.
From the IR point of view, sSQCD is reminiscent of the multitrace
deformation discussed in~\cite{Essig:2008kz}.  The main difference can be
found in the fermionic sector, where the goldstino also has a component in
the $\psi_S$ direction.  As such, multitrace deformations are not needed
here to give reasonable masses to the fermionic components of $X$.  The
phenomenology of sSQCD is thus very similar to the phenomenology
of~\cite{Essig:2008kz}.

At this point one may observe that there appears to be a contradiction
between our result \sSQCDapprox for $M_\text{gaugino}$, using also the
explicitly computed $g(x_\mathcal{M})$ of \fgsSQCD,  and the general result
of small first-order gaugino mass of Komargodski and Shih
\cite{Komargodski:2009jf}. However, this is not so: our example is in a
sense modular. The gaugino mass appears proportional to
$\vev{F_S}/\vev{S}$, for it arises from the extra SUSY-breaking sector we
have included. This can then be thought of as a separate hidden sector,
with ISS as the messenger sector. The treatment of Komargodski and Shih
does not constrain such models.

\newsec{Discussion and conclusion}[conc]
In this paper we have used the results of \cite{Fortin:2011ad} to further
illustrate how the OPE can be used to understand superpartner spectra in
the MSSM in the context of gauge mediation.  Although delivering only
approximate answers, our methods do capture the essential physics of
soft-mass generation in the MSSM. This becomes possible through the UV-IR
splitting achieved by the OPE.  The methods developed here lead to
approximations valid up to order one numbers both at weak and strong
coupling, as can be checked explicitly for strongly-coupled theories with
weakly-coupled duals. For strongly-coupled theories of SUSY breaking
without weakly-coupled duals, the logic can be inverted and the
approximations discussed here might allow us to argue for the functional
dependence of relevant anomalous dimensions, which are in practice
technically very difficult to calculate.

Using similar techniques one should also be able to perform approximate
computations of total cross-sections from the visible sector to the hidden
sector, which could be very useful in the event that SUSY is discovered at
the LHC.

Our methods were applied here to a new deformation of SQCD, where an
additional spontaneous breaking of SUSY is considered. This arises from the
F-term vev of a spurion $S$, whose zero component supplies the quark masses
in SQCD.  This deformation moves the ISS vacuum away from the origin and
thus induces a breaking of the accidental R-symmetry. Consequently,
Majorana gaugino masses are allowed in this ISS-like vacuum. Note that
there are no SUSY vacua with our deformation of SQCD. An obvious extension
of our work would be to study the $\mu/B_\mu$ problem in strongly-coupled
models, although this is bound to be more model-dependent.

In \MassApproxIR and \sSQCDapprox, the main results of this paper, the soft
masses are parametrized by entries of the anomalous-dimension matrix
$\gamma$ between the current $K$ and the spurionic operator $S^\dagger S$.
The calculation of $\gamma$ can be easily done in the UV, where the
electric theory is under control, with the one-loop result \AnomMatWeak.
One could then imagine using magnetic variables to express $\gamma$ in a
form useful in the IR, but the presence of the electric coupling $g$ in
\AnomMatWeak complicates matters. A direct calculation of $\gamma$ in the
IR of SQCD, around the SUSY-breaking minimum, using the magnetic
description from the outset, is thus more desirable.  However the meaning
of the current $K$ in terms of the magnetic dual fields is not clear
\textit{a priori}.  It would be interesting to carry out in mSQCD the
computations done for MGM, and then determine some of the relevant
anomalous dimensions of mSQCD operators.

Finally, the approximation \MassApproxIR for the gaugino masses in theories
with only dynamical SUSY breaking does not depend on any F-terms.
Corrections from the small explicit R-symmetry breaking, necessary to
generate non-vanishing Majorana gaugino masses, modify the approximation
\MassApproxIR, however the main contribution resides in the vev of the
MSSM-restricted mesonic superfield, which carries the appropriate R-charge.
For hidden-sector gauge groups that are completely Higgsed, this implies
that the vev of the MSSM-restricted mesonic superfield must either vanish
or blow up as the dynamical SUSY-breaking effect is taken to zero.  This
observation suggests that to obtain acceptable phenomenology, with
$|M_{\text{gaugino}}/m_{\text{sfermion}}|$ of order one, the spontaneous
R-symmetry breaking must be non-negligible.

\ack{We thank Antonio Amariti, Matthew Buican and Martin L\"{u}scher for
useful discussions. We are especially grateful to Ken Intriligator for
numerous helpful discussions and suggestions, and for comments on the
manuscript.  JFF is supported by the ERC grant BSMOXFORD No.\ 228169.  AS
is supported in part by the U.S.\ Department of Energy under contract No.\
DOE-FG03-97ER40546.}

\appendix{}
\newsec{Superpartner spectra in weakly-coupled theories}[app:Supspec]
Superpartner spectra can be computed directly from the messenger sector
in weakly-coupled theories of SUSY breaking.  Although the result is well-known
for simple messenger sectors, as for example in MGM~\cite{Martin:1996zb},
for general messenger sectors this is not the case (for a derivation using GGM,
see~\cite{Marques:2009yu}).

Consider a messenger sector consisting of $n$ chiral superfields $\Phi_i$ and
$\widetilde{\Phi}_i$ transforming in a vector-like representation $R+\bar{R}$
of the MSSM with arbitrary mass matrices $\left(\bbM_0^2\right)_{2n\times 2n}$
and $\left(\bbM_{1/2}\right)_{n\times n}=W_{ij}$ such that
\eqn{\mathscr{L}\supset-\begin{pmatrix}\phi_i^* & \tilde{\phi}_i\end{pmatrix}\left(\bbM_0^2\right)_{ij}\begin{pmatrix}\phi_j\\\tilde{\phi}_j^*\end{pmatrix}-\begin{pmatrix}\tilde{\psi}_i\end{pmatrix}\left(\bbM_{1/2}\right)_{ij}\begin{pmatrix}\psi_j\end{pmatrix}-\text{h.c.},}
where $(\phi_i,\psi_i)$ are the bosonic and fermionic components of $\Phi_i$
and similarly for $\widetilde{\Phi}_i$.  Introducing unitary matrices $U_b$, $U_f$
and $\tilde{U}_f$ which diagonalize the mass matrices,
\eqn{\tilde{m}_i^2\delta_{ij}=(U_b\,\bbM_0^2\,U_b^\dagger)_{ij},\qquad m_i\delta_{ij}=(\tilde{U}_f^*\,\bbM_{1/2}\,U_f^\dagger)_{ij},}
with $\tilde{m}_i$ and $m_i$ the (real positive) bosonic and fermionic mass
eigenvalues respectively, the gaugino and sfermion masses are given by
\eqn{M_{\text{gaugino}} =
-\frac{\alpha_\text{SM}}{\pi}\,C(R)\,\mathscr{G},\qquad
m_{\text{sfermion}}^2= \left(\frac{\alpha_\text{SM}}{4\pi}\right)^2\,C_2(R_\text{sfermion})\,C(R)\,\mathscr{F}^2,}[suppart]
where $C(R)=\tfrac{1}{2}$ for the fundamental representation and
\eqna{\mathscr{G} &= \sum_{i=1}^{2n}\sum_{j,k,l=1}^n(U_b)_{ik}(U_b^*)_{i,n+l}(U_f^\dagger)_{kj}(\tilde{U}_f^\dagger)_{lj}\,m_j\left[\ln\left(\frac{\Lambda^2}{m_j^2}\right)-\frac{\tilde{m}_i^2}{\tilde{m}_i^2-m_j^2}\ln\left(\frac{\tilde{m}_i^2}{m_j^2}\right)\right],\displaybreak[0]\\
\mathscr{F}^2 &= \sum_{i=1}^{2n}\tilde{m}_i^2\ln(\tilde{m}_i^2)[4+\ln(\tilde{m}_i^2)]+4\sum_{i=1}^nm_i^2\ln(m_i^2)[-2+\ln(m_i^2)]\\
 &\quad +\sum_{i,j,k,l=1}^{2n}(-1)^{\lfloor(k-1)/n\rfloor+\lfloor(l-1)/n\rfloor}(U_b)_{ik}(U_b^\dagger)_{kj}(U_b)_{jl}(U_b^\dagger)_{li}\displaybreak[0]\\
 &\quad \hspace{1cm}\times\,\tilde{m}_i^2\left[-\ln(\tilde{m}_j^2)\ln(\tilde{m}_j^2)+2\ln(\tilde{m}_i^2)\ln(\tilde{m}_j^2)-2\LiTwo\left(1-\frac{\tilde{m}_i^2}{\tilde{m}_j^2}\right)\right]\displaybreak[0]\\
 &\quad +2\sum_{i=1}^{2n}\sum_{j,k,l=1}^n\left[(U_b^\dagger)_{ki}(U_f)_{jk}(U_b)_{il}(U_f^\dagger)_{lj}+(U_b^\dagger)_{n+k,i}(\tilde{U}_f^*)_{jk}(U_b)_{i,n+l}(\tilde{U}_f^T)_{lj}\right]\displaybreak[0]\\
 &\quad \hspace{1cm}\times\left\{\tilde{m}_i^2\left[\ln(m_j^2)\ln(m_j^2)-2\ln(\tilde{m}_i^2)\ln(m_j^2)+2\LiTwo\left(1-\frac{\tilde{m}_i^2}{m_j^2}\right)-2\LiTwo\left(1-\frac{m_j^2}{\tilde{m}_i^2}\right)\right]\right.\displaybreak[0]\\
 &\quad \left.\hspace{1cm}+m_j^2\left[\ln(\tilde{m}_i^2)\ln(\tilde{m}_i^2)-2\ln(\tilde{m}_i^2)\ln(m_j^2)+
 2\LiTwo\left(1-\frac{\tilde{m}_i^2}{m_j^2}\right)+2\LiTwo\left(1-\frac{m_j^2}{\tilde{m}_i^2}\right)\right]\right\}.}
The diagrams leading to \suppart can be found in~\cite{Martin:1996zb}.  Note
that due to the magic of SUSY, the cutoff $\Lambda$ does not appear in the
gaugino masses.  Here $\LiTwo(x)=-\int_0^1\,dt\,\frac{\ln(1-xt)}{t}$ is
the dilogarithm or Spence function.

Note that, although the messenger spectrum of MGM, mSQCD and sSQCD are quite
different, the superpartner spectra are given in terms of the same functions
$g(x)$ and $f(x)$ (when the spontaneous R-symmetry breaking contribution is
discarded in the sSQCD case).

As a final point, note that it is straightforward to include extra messengers
transforming under different representations of the MSSM gauge group.

\bibliography{SCMGMRef}

\end{document}